\documentstyle[12pt,psfig]{article}
\def\bea{\begin{eqnarray}}
 \def\eea{\end{eqnarray}}
\begin{document}
\thispagestyle{empty}
\newcommand{\gsim}{ \mathop{}_{\textstyle \sim}^{\textstyle >} }
\newcommand{\lsim}{ \mathop{}_{\textstyle \sim}^{\textstyle <} }
\noindent
\begin{flushright}
        OHSTPY-HEP-T-98-012\\
        July 1998
\end{flushright}

\vspace{1cm}
\begin{center}
  \begin{Large}
  \begin{bf}
  The Phenomenology of SUSY models with a Gluino LSP
   \\
  \end{bf}
 \end{Large}
\end{center}
  \vspace{1cm}
 
    \begin{center}
    Stuart Raby$^\dagger$ and Kazuhiro Tobe$^*$\\
      \vspace{0.3cm}
\begin{it}
Department of Physics, \\
The Ohio State University, \\
174 W. 18th Ave., \\
Columbus, Ohio  43210\\
$^\dagger$raby@mps.ohio-state.edu \\ $^*$tobe@pacific.mps.ohio-state.edu\\
\end{it}

  \end{center}
  \vspace{1cm}
\centerline{\bf Abstract}
\begin{quotation}

In this paper we study the experimental consequences of a theory which
naturally has a heavy, stable (or almost stable) gluino.
We define the boundary conditions at a messenger scale $M \sim
10^{14}$ GeV which lead to this alternative phenomenologically acceptable 
version of the minimal supersymmetric standard model.  

In this theory, either the gluino or the gravitino is the lightest
supersymmetric particle [LSP].  If the gravitino is the LSP, then the gluino 
is the next-to-LSP with a lifetime on the order of 100 years.  In either case, 
the gluino is (for all practical purposes) a stable particle with respect to 
experiments at high energy accelerators.  Thus the standard missing energy
signature for SUSY fails.    A stable gluino forms a color singlet hadron, 
the lightest of which is assumed to be an isoscalar gluino-gluon bound state ($R_0$).  
The $R_0$ has strong interactions and will interact in a hadronic calorimeter; depositing
some fraction of its kinetic energy.   Finally, in the case the gravitino is the
LSP, bounds from searches for stable heavy isotopes of hydrogen or oxygen
do not apply to the metastable $R_0$.

\end{quotation}
\vfill\eject 

\section{Introduction}
 
Supersymmetry [SUSY] is a strongly motivated candidate for new physics
beyond the Standard Model [SM].  The minimal supersymmetric particle content is 
well defined and the interactions of all the new superparticles [sparticles]
are constrained by the observed SM interactions as long as the theory has a 
conserved R parity.  In order to search for SUSY one necessarily focuses on 
specific signatures which may rise above the standard model backgrounds.   
These signatures depend on how supersymmetry is broken; i.e. on the 
sparticle spectrum.  There are two classes of models for SUSY breaking which 
have been studied extensively. They are the minimal supergravity
model\cite{sugra} (or constrained minimal supersymmetric model
[CMSSM]\cite{cmssm}) defined by 5 parameters and low energy gauge-mediated 
SUSY breaking [GMSB]\cite{gmsb,dine} defined by 3 parameters.  
The 5 parameters in the CMSSM model are a universal scalar mass $m_0$, 
a universal gaugino mass $M_{1/2}$, a trilinear scalar coupling proportional 
to $A$ and $\mu, \; \mu B = m_3^2$ in the Higgs sector.  In this theory, 
the LSP is the lightest neutralino or a sneutrino.  The dominant signature 
for SUSY is missing energy due to the escape of the LSP from the detector.  
In low energy GMSB, the effective scale of SUSY breaking is given by 
$\Lambda = F/M \approx M \sim 10^5$ GeV.  The other two parameters are 
$\mu, \; \mu B = m_3^2$ as before.   In this case, the gravitino with
mass of order 10 - 1000 eV is the LSP.  The next to lightest SUSY particle
[NLSP], for example the lightest neutralino, can decay into a gravitino and a 
photon.   
This can lead to energetic photons plus missing energy, such as the single 
$e^+ e^- \gamma \gamma$ + missing energy event seen at CDF.

In a recent paper\cite{raby} it was shown how GMSB with Higgs-messenger mixing 
in an SO(10) theory naturally leads 
to a gluino LSP.  Since the gluino carries color, it will be
confined.  The stable LSP is then assumed to be a gluino-gluon bound state, a 
glueballino ($R_0$).  $R_0$ is a hadron and thus interacts strongly in a 
hadronic calorimeter.  As a result the standard missing energy signature for 
SUSY fails.  This theory has a characteristically different spectrum of squark, slepton 
and gaugino masses.   
In this paper we analyze the low energy spectrum of the theory, starting with boundary 
conditions at the messenger scale $M \sim 10^{14}$ GeV, using two (one) loop 
renormalization group equations for dimensionless (dimensionful) parameters 
down to the weak scale, $m_Z$.  We self-consistently require electroweak 
symmetry breaking using the one-loop improved Higgs potential.   The model has 
six SUSY breaking parameters --- $\Lambda \sim 10^5 \; \rm GeV, a \sim b 
\sim 10^{-1} - 10^{-2}, \; \mu, \; \mu B = m_3^2$ and $d$ (the size of the D term
contribution in units of $M_2$).  One of these parameters, $b$, allows us to 
arbitrarily vary the gluino mass.

Finally we note that the version of the model presented in 
this paper includes the SUSY breaking contribution
of an anomalous U(1)$_X$, in addition to GMSB at the messenger scale
$M$\cite{raby}.  
We show that the additional D term contribution is necessary in order to 
obtain a phenomenologically acceptable theory.  In particular,  without the 
D term the light Higgs boson $h$ would be {\em unacceptably} light.   
In the following, we assume that
the contribution to scalar masses from GMSB and the U(1)$_X$ D term are
comparable. In a recent paper\cite{tobe}, we 
have presented a model which dynamically breaks SUSY and leads to {\em
comparable} SUSY breaking effects from gauge-mediated and D term sources.  
The messenger scale in this theory is determined by the Fayet-Iliopoulos 
D term\cite{fidterm} of U(1)$_X$.\footnote{The magnitude of the Fayet-Iliopoulos
D term $\xi$ is given by $\xi   = \epsilon \; M_{Pl}^2 $
where $\epsilon  =  {g_X^2 Tr {\bf Q^X} \over 192 \; \pi^2} ({\sqrt{2} M_{st} 
\over g_X \; M_{Pl}})^2$ and the string scale $M_{st}$ is determined by
 the compactification scale.  In weak coupling $M_{st} = {g_X \over \sqrt{2}} \;
M_{Pl}$. In the strong coupling limit of the $E_8\times E_8$ heterotic 
string\cite{horava}, on the other hand, the compactification scale is 
identified with the GUT scale. Note, however, in a recent paper\cite{jmr} 
it has been argued that even in the strong coupling limit of the heterotic 
string, the scale of the D term is set by the Planck scale and not the 
compactification scale.  This argument has some caveats as discussed in the
paper.  For example, it might not apply if the anomalous gauge
symmetry eminates from D brane modes.  Another possibility is that the anomalous
U(1) in the effective four dimensional theory is a linear combination of U(1)s
on the two 10 dimensional boundaries of the $E_8\times E_8$ heterotic string.
In this case the axion which cancels the anomally is partially in the S and
T moduli of the theory.  As a consequence the scale of the anomally becomes
arbitrary.} This theory suggests that the boundary conditions we consider may
be ``natural."

\section{Boundary conditions at the Messenger scale}

The boundary conditions at the messenger scale are determined by two sources of
SUSY breaking --- gauge-mediated and D term. 

\subsection{Gauge mediated SUSY breaking}

The messenger sector is defined by an SO(10) invariant superspace potential
\begin{eqnarray}
W  \supset &  \lambda_a \;{\bf 10_H} \; A \; {\bf 10} + \lambda_s \; S \; {\bf 
10}^2 & \nonumber \\
 & + \lambda_1 \; \overline{\bf 16}_1 \; A \; {\bf 16}_1 +
\lambda_2 \; \overline{\bf 16}_2 \; A \; {\bf 16}_2 + \lambda \; S \; 
\overline{\bf 16}_1 \; {\bf 16}_2 &  
\end{eqnarray}
The adjoint $A$ is assumed to get a vacuum expectation value [vev]
\bea  <A> & = (B - L) \; M_G  & \eea
where $B - L$ (baryon number minus lepton number) is non-vanishing on color 
triplets and zero on weak doublets in the ${\bf 10}$ and the singlet $S$ 
is assumed to get a vev 
\bea 
<S(x,\theta)> & = S + \theta^2 \; F_S &.
\eea
We take $S \sim 10^{14}$ GeV with the effective SUSY breaking scale in the 
observable sector given by  
\bea
\Lambda & =  F_S/S  & \sim 10^5 \; \rm GeV.
\eea
The field  ${\bf 10_H}$ includes the weak doublet and color triplet Higgs, 
while ${\bf 10}$
contains the minimal set of messengers.   The first two terms in $W$ correspond to the
natural doublet-triplet splitting mechanism in SO(10)\cite{dw}.  Note that the doublet
messengers obtain mass at the scale $S$, while the triplet messengers obtain mass at the GUT
scale.  This theory also includes a natural mechanism for suppressing proton decay when  $S
<< M_G$\cite{babu}.  Finally, the last three terms in the superspace potential include four
additional sets of messengers.  They are introduced solely to give gluinos mass at one 
loop.\footnote{The first two terms do not contribute to gaugino masses at one loop due to an
accidental cancellation as shown by one of us (K.T.).}  Note, the R symmetry
which keeps gauginos massless is spontaneously broken at $S$.

In this theory, gauginos obtain mass at one loop
\bea
m_{\tilde g} & = {\alpha_3(M) \over  \pi} \; b^2 \; \Lambda & \nonumber \\
M_2 & = {\alpha_2(M) \over 4 \pi} \; \Lambda ( 1 + 4 b^2) & \nonumber \\
M_1  & = {3 \over 5} \,{\alpha_1(M) \over 4 \pi} \; \Lambda(1 + {20 \over 3} b^2) &
\label{eq:gaugino} \eea
where the parameter $b = \lambda \; S/ \sqrt{\lambda_1\, \lambda_2}\;M_G$
is derived from the last 3 terms in $W$.  The parameter $b$ is naturally of order
$10^{-2}\lambda/ \sqrt{\lambda_1\, \lambda_2}$.  For the purposes of our 
analysis, we take $b = 0.1$.  However $b$ is clearly a free parameter which 
may be used to vary the gluino mass.

Scalars obtain mass at two loops
\bea 
\tilde m^2 & = 2 \Lambda^2 \{ C_3 ({\alpha_3(M) \over 4 \pi})^2 (a^2 + 4 b^2) &
\nonumber \\
 & + C_2 ({\alpha_2(M) \over 4 \pi})^2 (1 + 4 b^2) &
\nonumber \\
& + C_1 ({\alpha_1(M) \over 4 \pi})^2 ({3 \over 5} + {2 \over 5} a^2 + 4 b^2)
\} &  \label{eq:scalar}
\eea
where $C_3 = 4/3$ for color triplets and zero otherwise, $C_2 = 3/4$ for weak
doublets and zero otherwise, $C_1 = {3 \over 5} (Y/2)^2$ and $a =\lambda_s S/
\lambda_a M_G$.  Note, the parameter  $a$ derives from the
first two terms in $W$. In our analysis, we take $a = 0.01$.  The
arbitrary parameters $a$ and $b$ are clearly independent.

Finally the gravitino mass is generically given by 
\bea m_{3/2} & =   F_{S}/\sqrt{3}\, M_{Pl} & \label{eq:m32} \eea   
with the reduced Planck scale, $M_{Pl} \equiv 1/\sqrt{8 \pi G_N} = 2.4
\times 10^{18}$ GeV.   Note, supergravity corrections 
to squark and slepton masses are proportional to the gravitino mass.

In the particular model of SUSY breaking discussed in \cite{tobe}, the field 
which gets both a scalar and F component vev is the
third component of an SU(2)$_F$ vector field,
$S_3$.  In this theory, the field $S$ is a composite superfield, with 
$S =  {S_3}^2/M_{st}$ and $F_S = 2 S_3 \; F_{S_3}/M_{st}$.  The scale $M_{st}$ is
associated with a string scale of order $M_G$.   Hence with a maximum value of 
$S_3 \sim 10^{15}$ GeV we obtain $S \sim 10^{-2}\; M_G \sim 10^{14}$ GeV.  
The upper bound on $S_3 \sim 10^{15}$ GeV has been assumed in order to suppress 
flavor changing neutral current processes induced by supergravity corrections 
to squark and slepton masses.   With the gravitino mass now given
by\footnote{The gravitino mass is set by the largest SUSY breaking vev in the
theory.   The D term contribution is of order the weak scale and thus is
negligible.   In the particular SUSY breaking sector of the theory given in 
\cite{tobe}, the fundamental SUSY breaking vev is given by
 $F_{S_3} = {M_{st} \over 2\, S_3} \; F_S \sim 10 \; F_S$.  This explains the change
in going from eqn. \ref{eq:m32} to \ref{eq:m322}.  It is clear that the 
value of the gravitino mass is model dependent.}
\bea m_{3/2} & =   F_{S_3}/\sqrt{3}\, M_{Pl} ,& \label{eq:m322}\eea   the ratio of the
supergravity contribution to squark and slepton masses (fixed by the gravitino mass) to
the GMSB contribution scales as
\bea m_{3/2}/M_2  & = {2 \pi \over \sqrt{3} \; \alpha_2} \; (S_3/M_{Pl}) \sim  10^2
(S_3/M_{Pl}) & \sim 0.04 .\eea  Hence with $S_3  \le 10^{15}$ GeV, supergravity gives at 
most a 0.16\%  correction to squark and slepton masses squared.  This may be sufficient to suppress 
large flavor changing neutral current processes. 

As an example of the spectrum at $M$ we have (for $\Lambda = 10^5$ GeV)
\bea
m_{\tilde g}  & = 13 \; (b/.1)^2 \; \rm GeV & \nonumber \\
M_2 & = 330 \; \rm GeV & \nonumber \\
m_{3/2} & = 12 \; (S_3/10^{15} \rm GeV)  \; \rm GeV .&  \label{eq:spectrum}
\eea
In addition, scalar masses are fixed by the largest SUSY breaking scale in the problem,
$M_2$; with right-handed squarks and sleptons being the lightest scalars obtaining mass
only via weak hypercharge interactions.

Thus, in this model the gluino (or possibly the gravitino) is naturally the 
LSP.   The gluino mass is suppressed by two powers of $S/M_G$.   One power 
comes because the color triplet messengers have mass of order $M_G$  and the 
second comes because the R symmetry
in this sector of the theory is only broken at the lower messenger scale $S$.

\subsection{D term SUSY breaking}

For phenomenological reasons we assume that SUSY is also broken by the
D term of an anomalous U(1)$_X$ gauge symmetry.   In a recent paper\cite{tobe}
we have shown that it is possible to obtain D term SUSY breaking and GMSB with
contributions to scalar masses of the same order.  In this paper we
simply parametrize the D term contribution to scalar masses by the formula
\bea 
\delta_D \tilde m_a^2 & =  Q^X_a \; d \; {M_2}^2 & \label{eq:dterm}
\eea
where $Q^X_a$ is the U(1)$_X$ charge of the field labelled by the
index $a$ and $d$ is an arbitrary parameter of order 1 which measures the
relative strength of D term vs. gauge mediated SUSY breaking.

Since we are working in the context of an SO(10) GUT,  $Q^X$ necessarily
commutes with the SO(10) generators.  In order not to consider a random
U(1) gauge symmetry we shall assume that U(1)$_X$ is embedded into E$_6$
with 
\bea
Q^X_a  & =   1    \hspace{.5in}   -2   \hspace{.5in}  4   & \nonumber \\
 \rm on   &  16  \hspace{.5in} 10  \hspace{.5in}  1  & \subset 27 \; \rm of \; E_6.
 \eea
This U(1)$_X$ is naturally family independent and is thus safe from inducing
large flavor changing neutral current processes.  It is also well-motivated in
the context of string theories\cite{farragi}.

\subsection{Summary of boundary conditions at $M$}

Thus the boundary conditions at the messenger scale $M$ 
from GMSB (for scalar and gaugino masses)
and from D term SUSY breaking (for scalar masses only) are given in
eqns. \ref{eq:gaugino}, \ref{eq:scalar}, and \ref{eq:dterm}.
There are 3 additional  parameters  ---  
the supersymmetric Higgs mass $\mu$, and the two soft SUSY breaking
parameters, i.e.  a scalar Higgs mass ${m_3}^2$ and the trilinear scalar coupling $A$.
To leading order we have \cite{giudice}
\bea  A & = 0 &  .\label{eq:a} \eea
  On the other hand, for the purpose of this analysis, 
we leave $\mu$ and ${m_3}^2$ arbitrary.  In principle, these parameters must 
be determined by new physics which we have not considered in this analysis.   
Using the 6 parameters  $\Lambda, \; a, \; b, \; d, \; \mu, \; 
{m_3}^2$  defining the boundary conditions at $M = 10^{14}$ GeV (eqns.
\ref{eq:gaugino}, \ref{eq:scalar}, \ref{eq:dterm} and \ref{eq:a};  $\mu$ and
${m_3}^2$ arbitrary)
we renormalize the full set of dimensionful (dimensionless) parameters at one 
(two) loop down to $M_Z$.  $\mu$ and $m_3^2$ are fixed by requiring electroweak
symmetry breaking.  For the bottom mass, we have included three loop QCD running
of $m_b(\mu)$ from  $\mu = m_b$ to $m_Z$ and one loop SUSY threshold corrections
at $m_Z$, relevant for large $\tan\beta$\cite{hall}.

\section{Sparticle masses at the Z scale}

 \subsection{Why introduce D term SUSY breaking?}

Consider first the low energy spectrum in the absence of D term scalar masses, 
i.e. $ d = 0$.  
%
\begin{figure}
	\centerline{\psfig{file=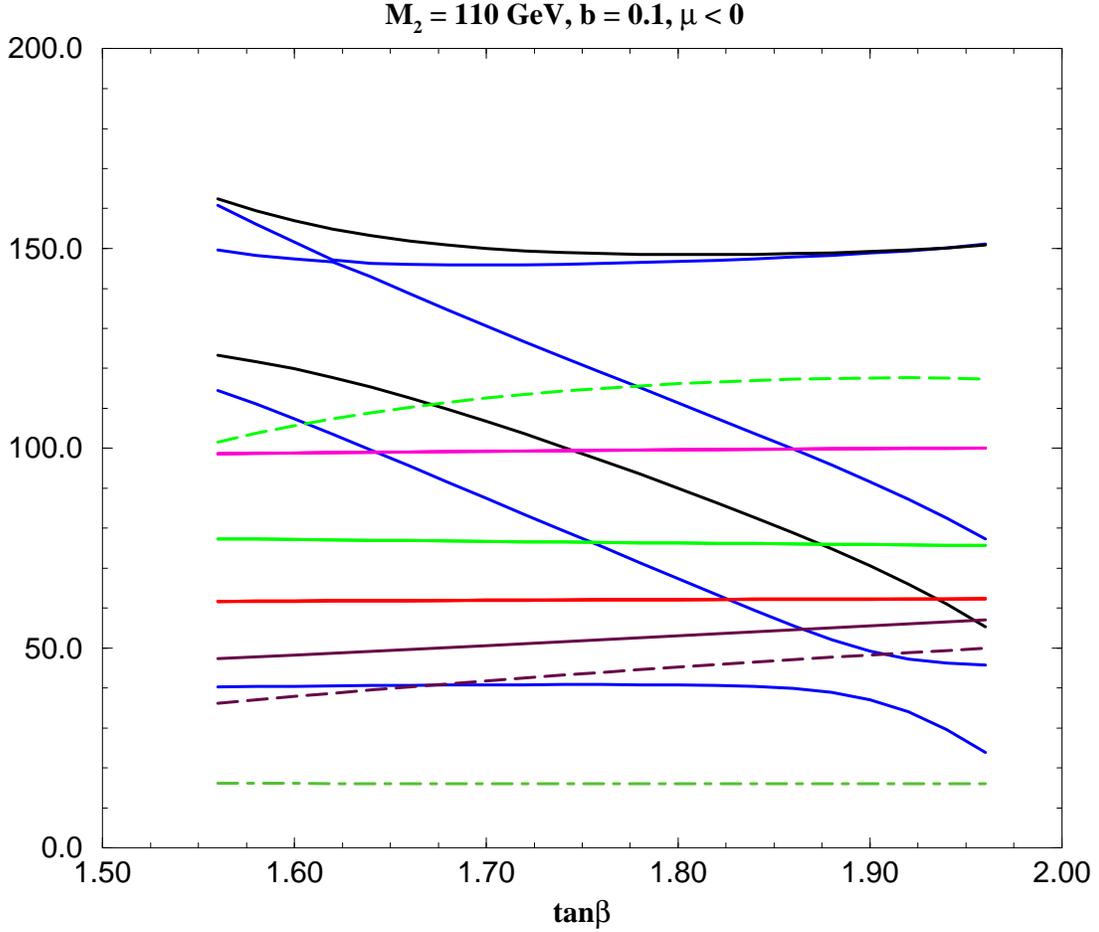,width=17cm,angle=-90}}
\caption{Mass spectrum of squarks, sleptons, neutralinos, charginos,
gluino, and Higgs for the case $M_2 = 110$ GeV, $\mu < 0, a=0.01$, 
and $b= 0.1$ plotted as a function of $\tan\beta$.  The right-handed up 
squarks (green), down squarks (red), sleptons (pink), neutralinos (blue), 
charginos (black), and gluino (dot-dashed green) masses are given by the 
respective colored lines. 
The first and second (third) generation squark and slepton 
masses are represented by solid (dashed) lines. 
The lightest Higgs mass is given by the solid (dashed) brown line at 
one loop level (tree level). }
\label{fig1}
\end{figure}
In fig. 1, we show the squark, slepton, Higgs and gaugino 
masses for the case $M_2 = 110 \; \rm GeV, \; \mu < 0, \; a = 0.01, \; \rm 
and \; b = 0.1$ plotted as a function of $\tan\beta$.  Note, the one-loop 
corrected light Higgs mass (the solid brown line) is below the present 
experimental lower bound,  $m_h > 90$ GeV\cite{Higgsbound}.  
We find $m_h < 60$ GeV.  This is due to the fact that at tree level the 
Higgs mass satisfies the approximate mass formula \bea  m_h & \sim m_Z \; 
\cos2\beta &  \eea with $\tan\beta \le 2$ in fig. 1.    
We could in principle increase the Higgs 
mass by increasing $\tan\beta$, however, due to the relation
\bea \mu^2 & = -{ m_Z^2 \over 2} + {m_{H_d}^2 - m_{H_u}^2 \,\tan^2\beta \over
\tan^2\beta - 1} & \eea as $\tan\beta$ increases,  $\mu$ decreases and as a
consequence the chargino mass also decreases.  The lower bound on the chargino 
mass $m_{\tilde \chi^+} \ge 85$ GeV\cite{charginobound} is violated for
$\tan\beta > 1.83$.\footnote{This bound assumes the standard missing energy 
signature for SUSY.  This limit is not applicable in our case.   However, 
there will be a lower bound on the chargino mass coming from the visible Z 
width or $e^+ e^- \rightarrow$ hadrons which  is comparable. We will discuss 
these limits on the chargino mass within our framework shortly.} Of course, 
we could in principle increase the chargino mass by increasing $M_2$, however, 
the right-handed SUSY breaking stop mass squared is negative at the weak scale 
and is driven further negative when $M_2$ increases since this is the dominant 
source of SUSY breaking.  For $M_2 \ge 150$ GeV,  the lightest stop mass 
becomes less than $m_Z/2$  (see fig. 2).  
\begin{figure}
	\centerline{\psfig{file=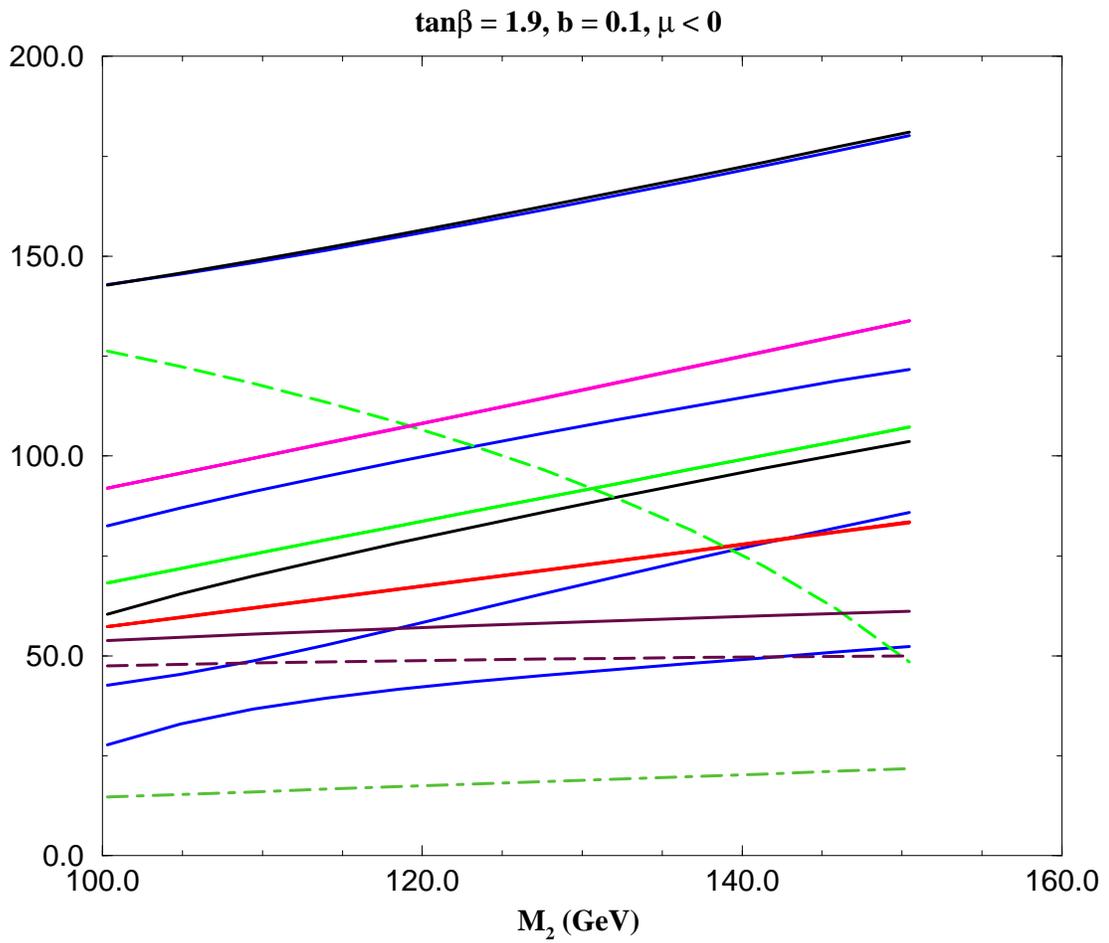,width=17cm,angle=-90}}
\caption{Mass spectrum of squarks, sleptons, neutralinos, charginos,
gluino, and Higgs for the case $\tan\beta=1.9, \mu <0$, and $b=0.1$ as a 
function of the wino mass $(M_2)$. The line notation is the same as in fig.1.}
\label{fig2}
\end{figure}
Thus we cannot make $M_2$ very
large.  Hence since both $\tan\beta$ and $M_2$ are severely constrained from 
above, the light Higgs mass is always below the experimental bound.  We have 
checked that the problem of a light Higgs, chargino and stop is even worse 
for $\mu > 0$.  

In order to solve this problem we have been forced to consider additional soft 
SUSY breaking mass contributions.   In order not to lose predictability, we 
have considered the well motivated addition of an anomalous U(1)$_X$ and its 
associated D term.  We now consider how this one new contribution resolves 
the light Higgs problem.

\subsection{The addition of D term scalar mass corrections}

As discussed, we parametrize the D term contribution to soft SUSY breaking 
scalar masses at the messenger scale by
\bea \delta_D \, \tilde m_a^2 & = Q^X_a \; d \; {M_2}^2  &  \eea with
\begin{equation}  Q^X = 1 \end{equation}
 for squarks and  sleptons and 
\begin{equation}  Q^X = -2 \end{equation}
for Higgs doublets.   Thus squarks and sleptons obtain a universal positive 
mass squared correction.  As a result we can now increase $M_2$ without the 
light stop mass becoming too small.   Then  increasing $M_2$ causes the 
chargino mass $m_{\tilde \chi^+}$ to increase.    Finally  $\tan\beta$ can now 
be increased which raises the light Higgs mass above the experimental lower 
bound. 

\renewcommand{\thefigure}{3 (\alph{figure})}\setcounter{figure}{0}
\begin{figure}
	\centerline{\psfig{file=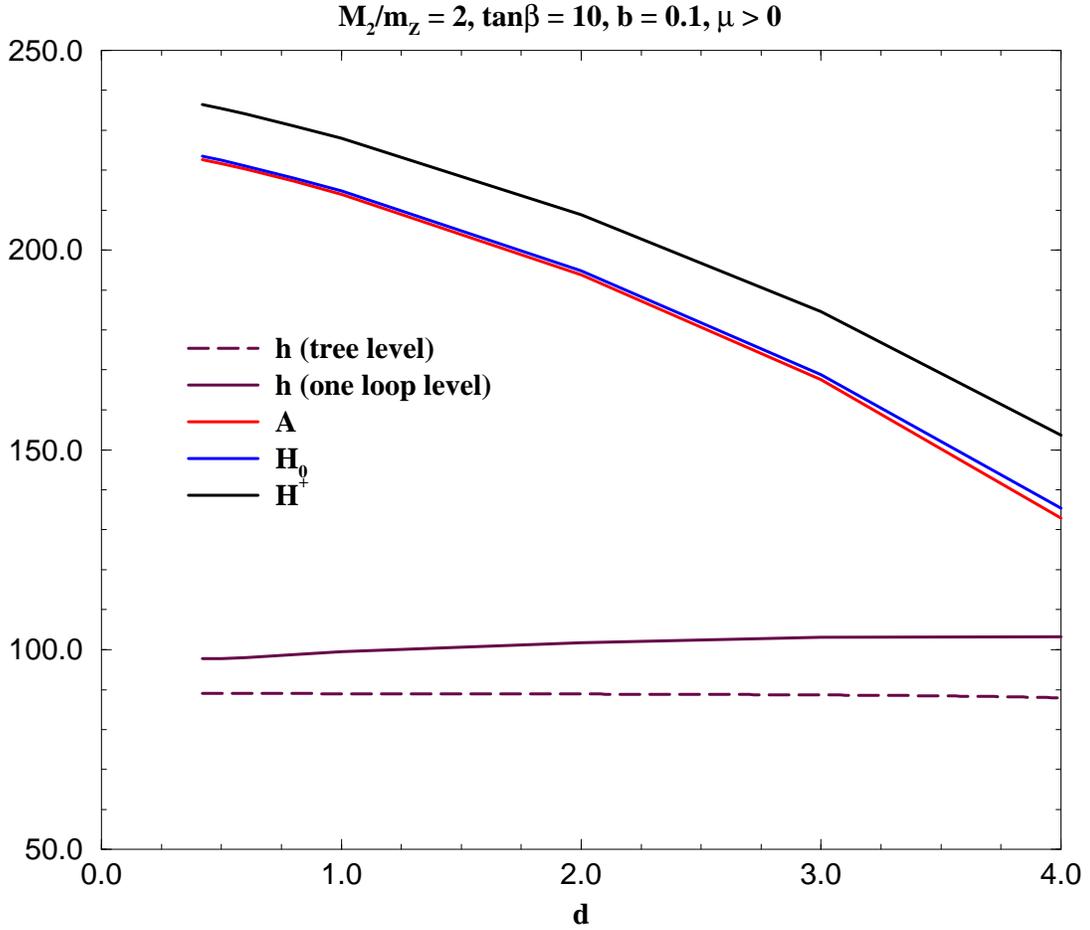,width=17cm,rheight=15.5cm,angle=-90}}
\caption{Mass spectrum of the light neutral Higgs $h$ (solid (dashed) 
brown) at one loop level (tree level), pseudoscalar boson $(A)$ (red), 
heavy neutral Higgs $(H_0)$ (blue), and charged Higgs $(H^+)$ (black)
for the case $M_2/m_Z = 2, \tan\beta=10, b=0.1$ and 
$\mu >0$ as a function of $d$.}
\label{fig3a}
\end{figure}
\begin{figure}
	\centerline{\psfig{file=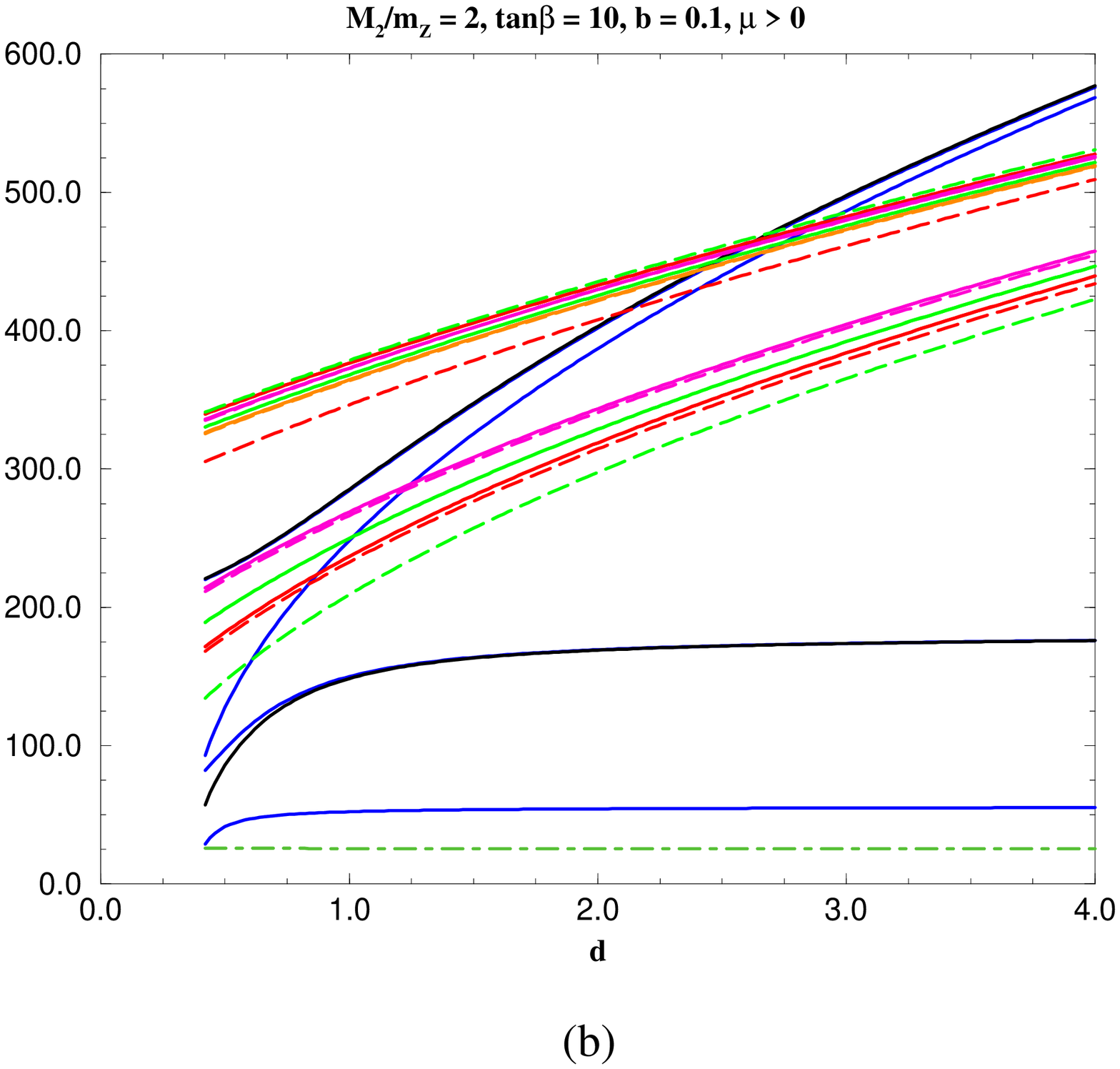,width=17cm,rheight=15.5cm,angle=-90}}
\caption{Mass spectrum of up squarks (green), down squarks (red),
sleptons (pink), sneutrinos (orange), neutralinos (blue), charginos (black),
and gluino (dot-dashed green) for the case $M_2/m_Z = 2, \tan\beta=10, b=0.1$ 
and $\mu >0$ as a function of $d$. 
The first and second (third) generation squark and slepton
masses are given by solid (dashed) lines.
}
\label{fig3b}
\end{figure}
In figs. 3(a,b) we plot the sparticle spectrum for $\tan\beta = 10$, 
$M_2/m_Z = 2$ 
as a function of $d$.  We see in fig. 3a that $98 \lsim m_h \lsim 103$ GeV 
for  $ 0.42 \lsim d \lsim 4.0$ (the lightest Higgs 
mass is given by the solid (dashed) brown line at one loop (tree level).  
Thus the light Higgs problem is solved with the addition of one new parameter, 
$d$.  The Higgs mass is determined by electroweak symmetry breaking and is 
thus only weakly dependent on $d$.  We also see that the mass of the other 
Higgs states, $A,\; H_0, \; H^+$, (evaluated at tree level) decrease as $d$ increases, with the 
pseudo-scalar $A$ remaining the lightest and for $d \lsim 3.5$, we have 
$m_A \gsim 150$ GeV.  Squark and slepton masses (fig. 3b), on the other hand, 
increase with $d$.   The light first and second generation squarks and  
sleptons are right handed; the heavy ones are left handed.  The third generation 
squarks and slepton eigenstates are  mixtures of left and right
handed states due to $A$ term mixing.  The lightest squark is either a stop
or sbottom.   Note, for $d \le 0.7$ the light
stop is lighter than the top.   This is  because the right handed stop mass squared is driven
negative by renormalization group running as a consequence of the large top yukawa coupling. 
This is the  same effect which drives the Higgs mass squared negative.  
Finally chargino and neutralino masses implicitly depend on $d$ through their 
dependence on $\mu$.   For small $d$ the lightest neutralino and chargino 
masses are naturally small.  The lower bound on $d$ is obtained by requiring
that the visible width of the Z be consistent with the experimental bound 
(see next section).   Recall, the value of the gluino mass depends on $b$.  
In the figures we have taken $b = 0.1$; the gluino mass scales as $b^2$. 

\renewcommand{\thefigure}{4 (\alph{figure})}\setcounter{figure}{0}
\begin{figure}
	\centerline{\psfig{file=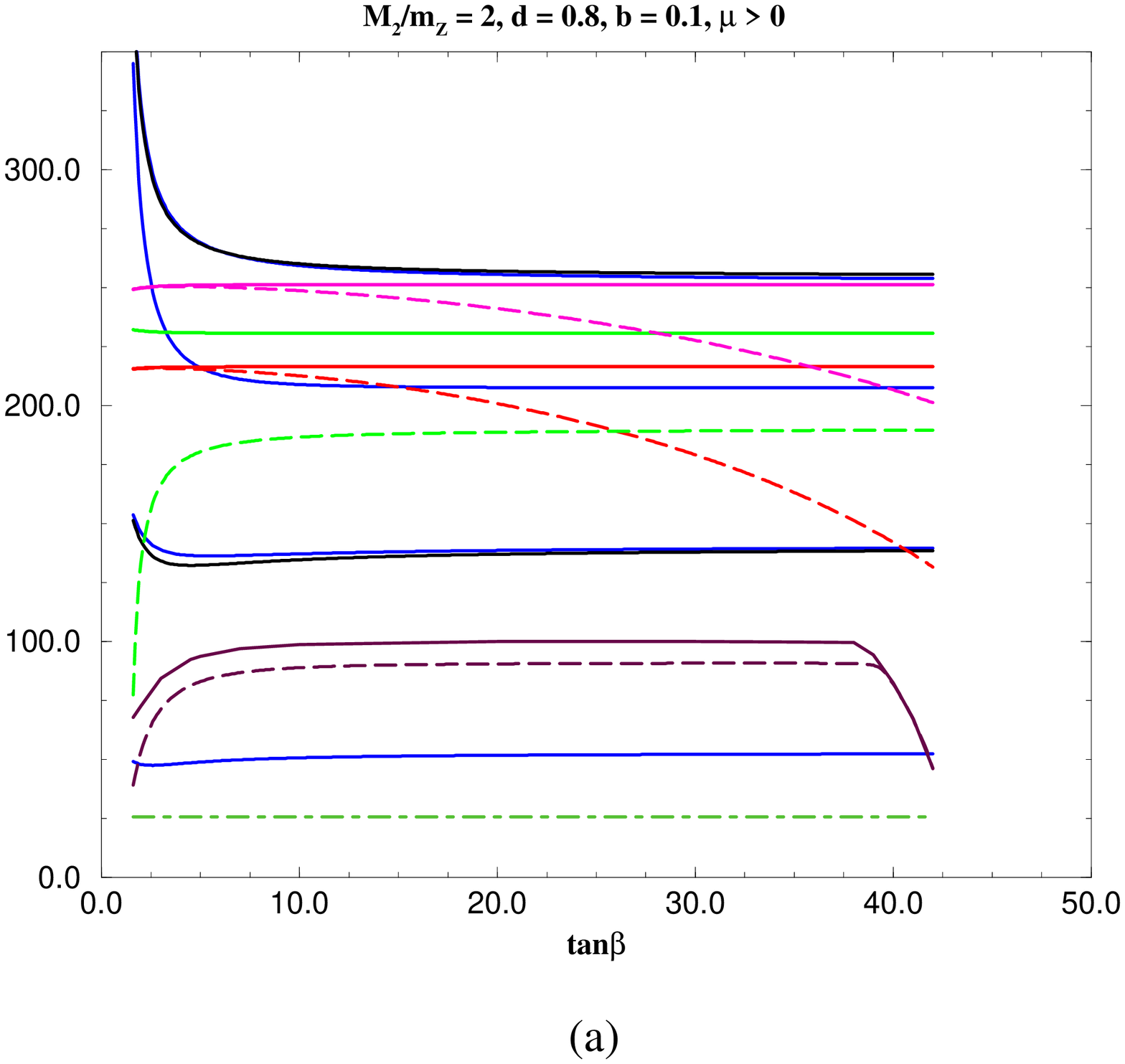,width=17cm,rheight=15.5cm,angle=-90}}
\caption{Mass spectrum of squarks, sleptons, neutralinos, charginos,
gluino, and Higgses for the case $d=0.8$, 
$M_2/m_Z = 2, b=0.1$, and $\mu >0$ as a function of $\tan\beta$.
The line notation is the same as in fig.1.}
\label{fig4a}
\end{figure}
\begin{figure}
	\centerline{\psfig{file=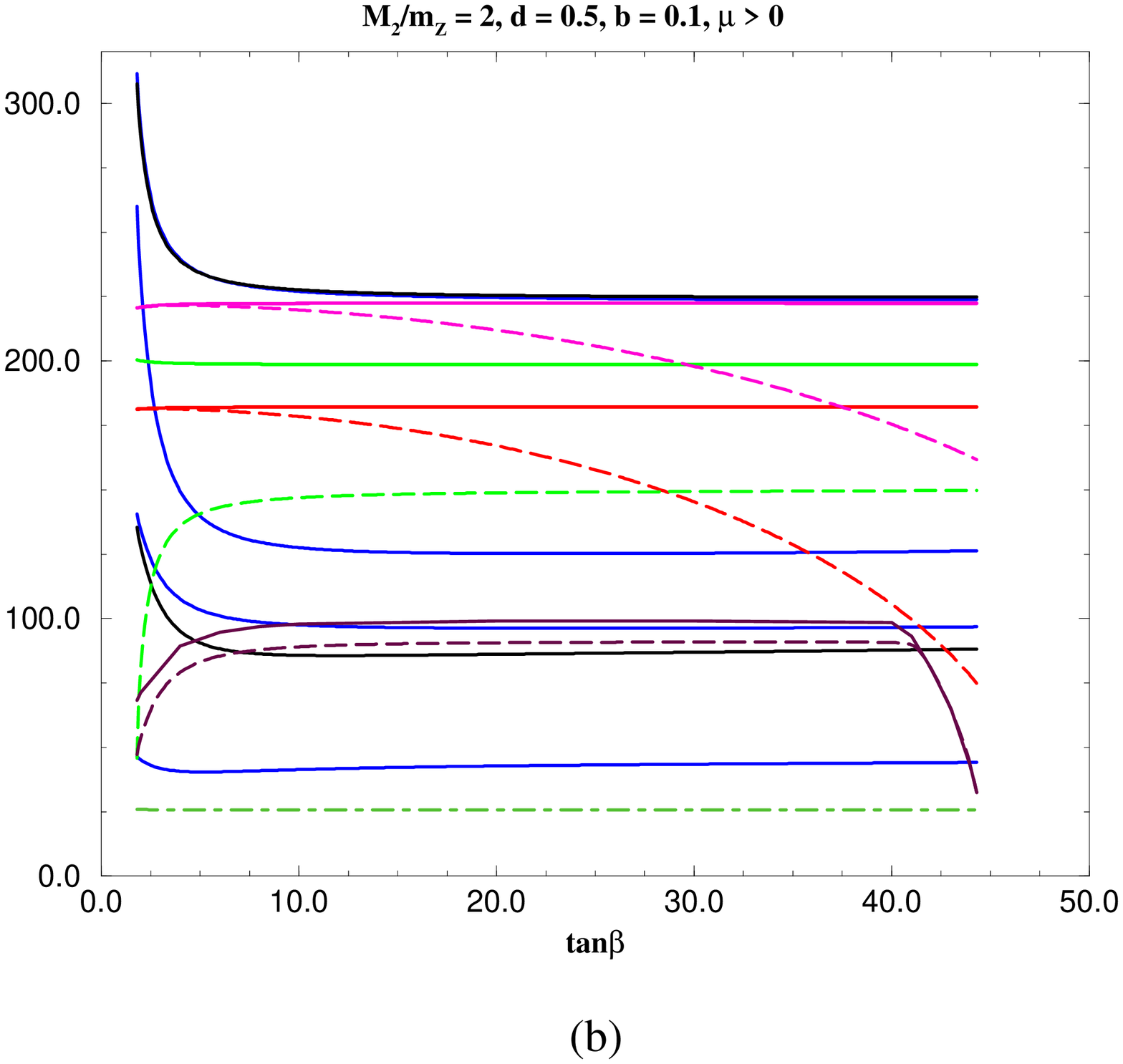,width=17cm,rheight=15.5cm,angle=-90}}
\caption{Same as fig.4 (a) except for $d=0.5$.}
\label{fig4b}
\end{figure}
\begin{figure}
	\centerline{\psfig{file=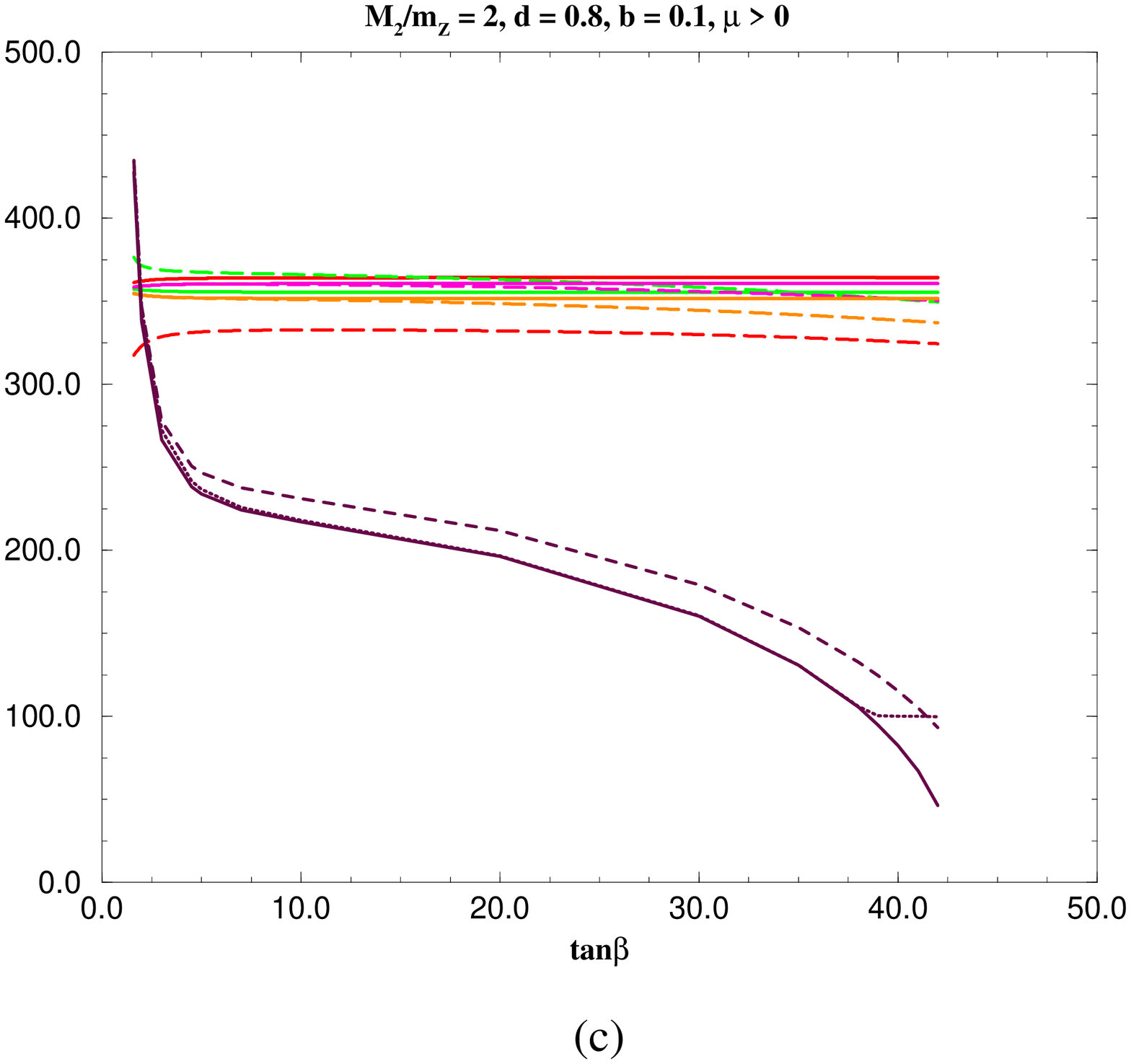,width=17cm,rheight=15.5cm,angle=-90}}
\caption{The mass spectrum of the heavy states (left-handed states) of the 
squarks and slepton (the line notation is same as in fig.3(b)), and 
peudoscalar boson (solid brown line), heavy neutral Higgs (dotted brown), 
and charged Higgs (dashed brown)
for the case $M_2/m_Z = 2, d=0.8, b=0.1$, and $\mu >0$ as a function of 
$\tan \beta$.}
\label{fig4c}
\end{figure}
In figs. 4, 5 we show the sparticle spectrum as a function of $\tan\beta$ and 
$M_2$ with the other parameters fixed.  In figs. 4(a,b) we take $M_2/m_Z = 2$, 
$d = (0.8, 0.5)$, $b = 0.1$ and $\mu > 0$.  We see (fig. 4a) that for 
$4 \lsim \tan\beta \lsim 39$ the light Higgs mass bound is satisfied.  For low 
and moderate values of $\tan\beta$ the pseudo-scalar ($A$) and charged Higgs 
($H^+$) are significantly heavier than $m_Z$.   Hence the phenomenology of the 
light Higgs is identical to that of the SM.  However for large $\tan\beta \sim 
30$ the bottom yukawa coupling is relevant.  In this case, the pseudo-scalar
mass $m_A^2 = m_{H_u}^2 + m_{H_d}^2$ becomes small (fig. 4c).  The upper bound on 
$\tan\beta$ is of order 38, since in this case both $m_h$ and $m_A$ are greater
than $\sim$ 75 GeV, the present experimental bound\cite{Higgsbound}.  
\renewcommand{\thefigure}{5 (\alph{figure})}\setcounter{figure}{0}
\begin{figure}
	\centerline{\psfig{file=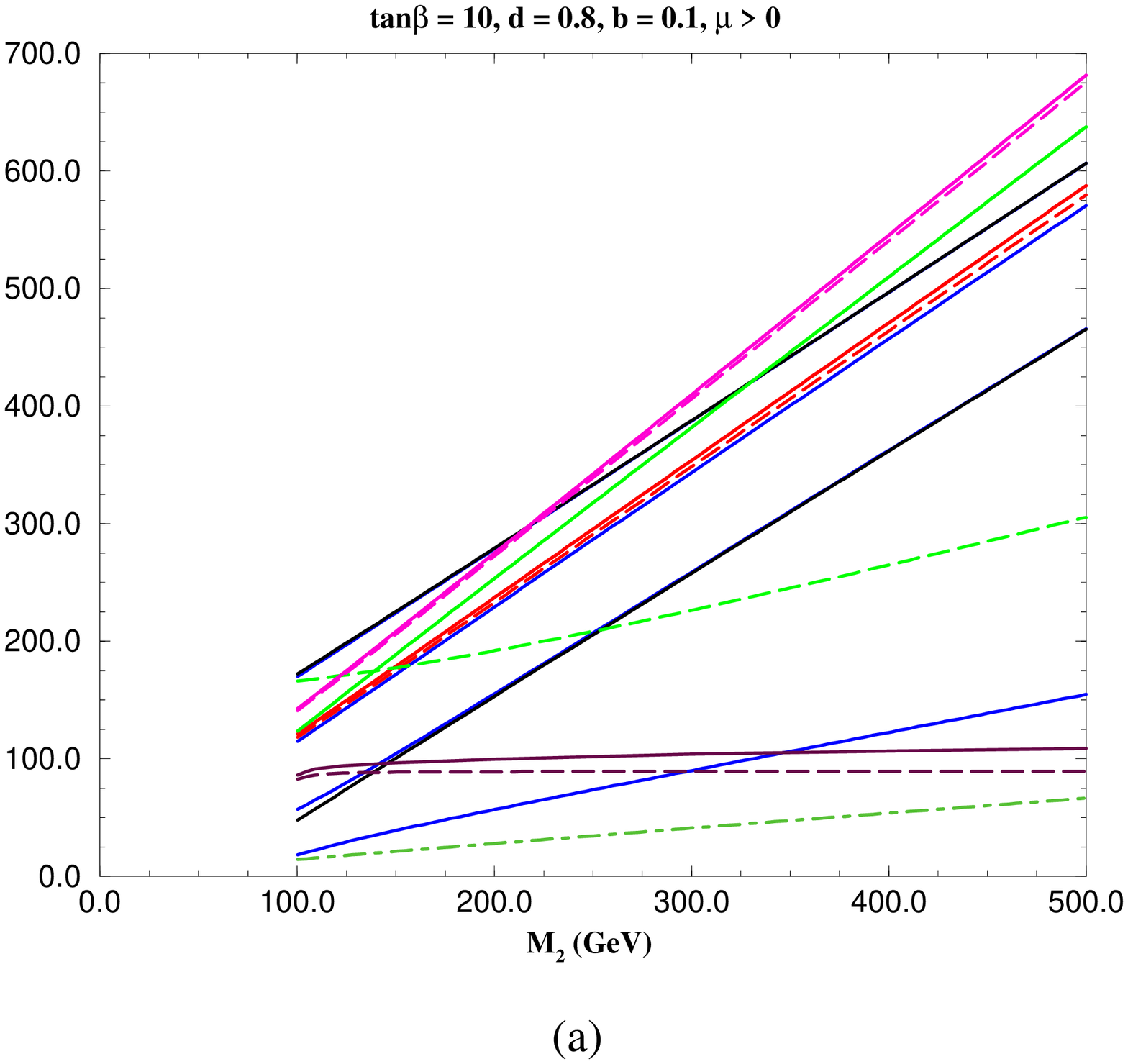,width=17cm,rheight=15.5cm,angle=-90}}
\caption{Mass spectrum of squarks, sleptons, neutralinos, charginos,
gluino, and Higgses for the case $\tan\beta=10$, $d=0.8, b=0.1$, and 
$\mu >0$ as a function of $M_2$. The line notation is the same as in fig.1.}
\label{fig5a}
\end{figure}
\begin{figure}
	\centerline{\psfig{file=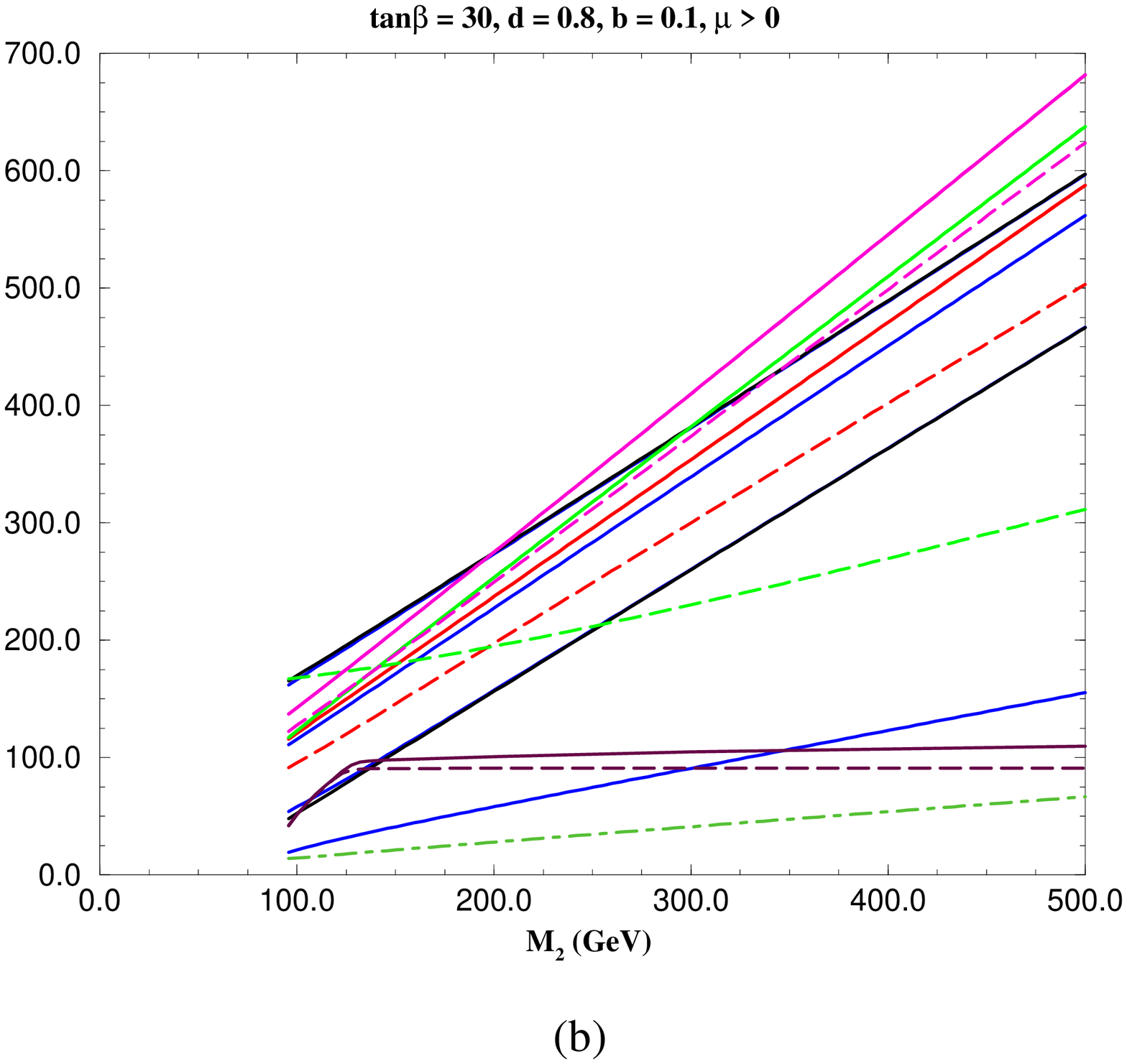,width=17cm,rheight=15.5cm,angle=-90}}
\caption{Same as fig.5 (a) except for $\tan\beta=30$.}
\label{fig5b}
\end{figure}
In figs. 5(a,b) we take $\tan\beta = (10, 30)$, $d = 0.8$, 
$b = 0.1$ and $\mu > 0$.  We see that the light Higgs mass increases slowly 
as $M_2$ increases with $m_h \lsim 115$ GeV. Squark, slepton and gaugino 
masses also 
increase as $M_2$ increases.  This is because all SUSY breaking masses scale 
with $\Lambda$.  For low values of $M_2$ of order 100 GeV, the lightest 
neutralino and chargino masses become small; the lower bound on $M_2$ is again 
determined by the visible width of the Z.

The gluino mass is arbitrary.  Its value is bound from below by about 1 GeV
since, in addition to the GMSB contribution to its mass (eqn. \ref{eq:gaugino}),
it receives a dynamical QCD contribution to its mass and also radiative 
corrections at the weak scale dominated by the stop; both of order a GeV.  
The gluino mass also increases as $M_2$ increases and for very large $M_2$ of 
order a TeV (fig. 6) the gluino mass is above 100 GeV.    
\renewcommand{\thefigure}{\arabic{figure}}\setcounter{figure}{5}
\begin{figure}
	\centerline{\psfig{file=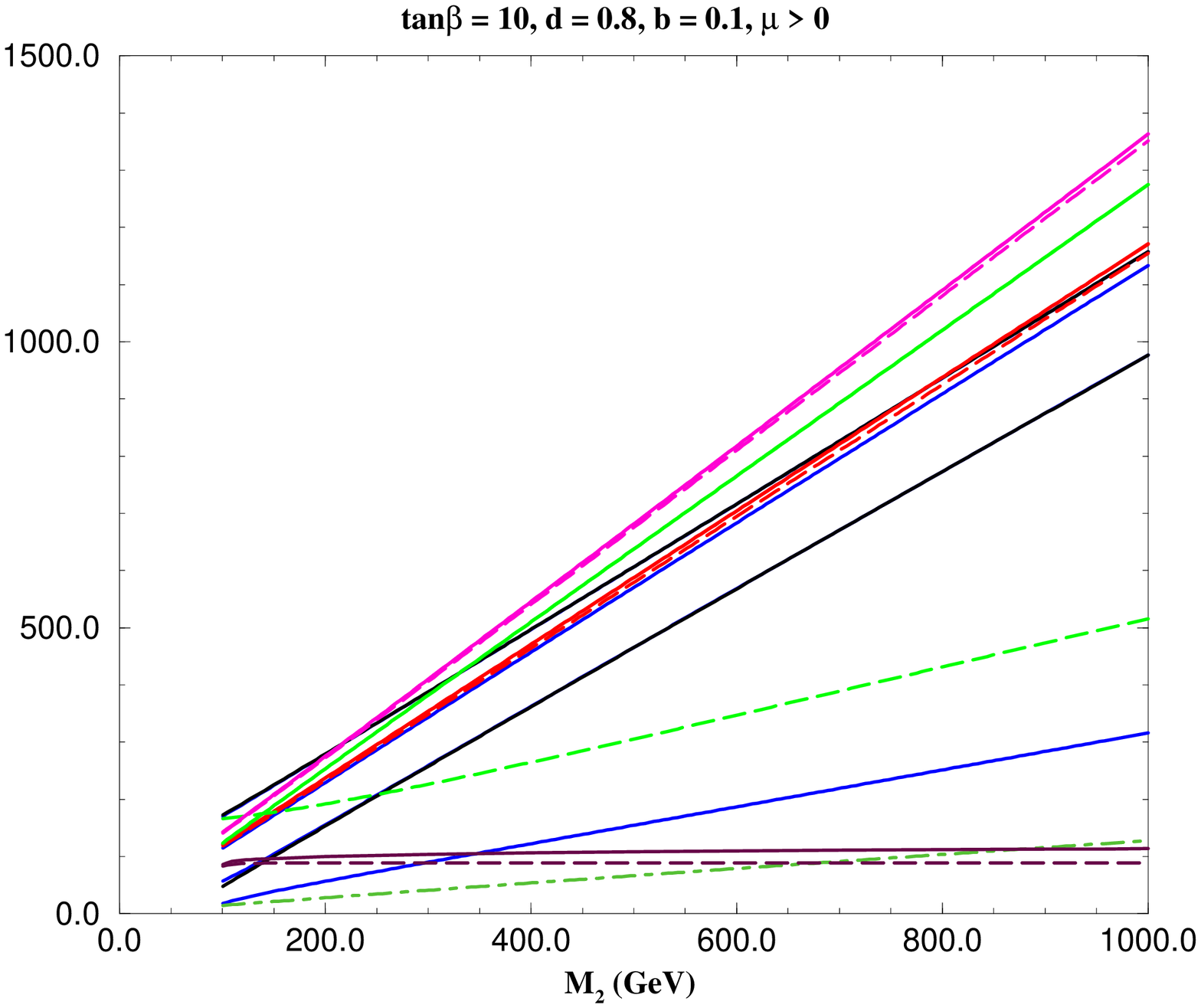,width=17cm,rheight=15.5cm,angle=-90}}
\caption{Same as fig.{\ref{fig5a}}}
\label{fig6}
\end{figure}

For a large range of the parameter $b$, either the gluino or the gravitino 
is the LSP.\footnote{Clearly if $b$ is increased, we could imagine a case in 
which the gluino is heavier than the lightest neutralino which is also 
somewhat light, particularly for small values of $M_2$ of order 100 GeV.  
In this case the neutralino is the LSP (or NLSP).  The gluino will then 
decay into the lightest neutralino.  This case is highly constrained by 
standard missing energy searches for gluinos.  However if the neutralino-gluino mass 
difference is only a few GeV, then the missing energy signal is suppressed and 
it may not be ruled out.  This possibility requires further study.  If,
in addition, the neutralino is also the LSP (lighter than the gravitino) 
it would be stable and would make an excellent dark matter candidate.  We do 
not consider this possibility further in this paper.}  In the case that the 
gravitino is the LSP, then the gluino is unstable with the decay rate for a 
gluino decaying into a goldstino ($G$) and a
gluon given by 
\bea  \Gamma(\tilde g \rightarrow G + g) & = {m_{\tilde g}^5 \over 16 \pi
F_{S_3}^2}\; \left( 1 - {m_G^2 \over m_{\tilde g}^2} \right)^3 &  \eea

or
\bea c \tau_{\tilde g} & \sim 10^{20} ({25 \rm GeV \over m_{\tilde g}})^5 ({\sqrt{F_{S_3}}
\over 6\times 10^9 \rm GeV})^4   cm & \label{eq:ctau} \eea
where in eqn. \ref{eq:ctau} we have neglected the gravitino mass.
{\em Hence for all laboratory experiments the gluino may be considered stable, 
decaying outside the detector.}  Thus the standard missing energy signature for
SUSY is gone.   The gluino will hadronize; forming a hadronic jet.  For a heavy
gluino, the jet may contain only the lightest stable hadron containing the
gluino.   In any event, some fraction of the gluino kinetic energy will be
visible in a hadronic calorimeter.

We assume that the LSP is a gluino-gluon bound state, a glueballino
($R_0$).\footnote{The other likely candidate LSP is a gluino--($u \, \bar u - d
\, \bar d)_8$ bound state, the neutral component of an isotriplet $\tilde \rho$.
We shall not consider this possibility further in this paper, see for eg. 
\cite{raby}.}  It is (for all practical purposes) stable because of a conserved R parity.  $R_0$ will 
interact in a hadronic calorimeter with a strong interaction cross-section.  
The dominant process at small momentum transfer is governed
by Regge exchange.  In a recent paper by Baer et al.\cite{baer} an estimate 
for the energy loss of $R_0$ in a hadronic calorimeter was obtained using the 
triple pomeron amplitude for the single diffractive process 
$R_0 \;N \rightarrow R_0 \; X$ where $X$ denotes the inclusive sum over all 
final states.  It was found that a 25 GeV gluino with $\beta \sim 1$ would 
deposit less than 20\% of its kinetic energy in the detector assuming 8 
hadronic collisions.   An $R_0$, however, could also have a significant charge 
exchange cross-section, given by a triple- Reggeon amplitude for the single 
diffractive process $R_0 \;N \rightarrow \tilde \rho^+ \; X$.  In this case 
there would be an additional ionization energy loss in the detector, assuming
$\tilde \rho^+$ is sufficiently long lived.   Baer et al. have assumed that at each
hadronic collision the light brown muck surrounding the gluino is stripped off and then
the gluino re-hadronizes.  They parametrize the probability for the resulting
gluino bound state to be a charged $\tilde \rho^+$ by $P$.
We shall discuss their results in the next section.  It suffices now to say that
their results are sensitive to this universal parameter $P$.  
Clearly a better theoretical estimate of $R_0 \, N$ scattering is needed.  
It is also clear that detailed detector 
simulations are necessary in order to confirm their results.
  
We can now test the theory with regards to LEP, CLEO and Tevatron data.  
 Note that flavor 
changing neutral current processes are naturally suppressed within our 
framework of combined gauge-mediated and D term SUSY breaking.

\subsection{Experimental tests}

First let's consider some tests from LEP.  Since the LSP in this theory is a 
gluino, charginos and neutralinos can now decay into $q \bar q \tilde g$.  
Moreover since charginos and neutralinos are relatively light they can be 
produced at relatively low center of momentum energies at LEP.  We have 
considered limits from two processes --- the visible width of the Z ($\Delta 
\Gamma_Z(\rm visible)$) and $e^+ e^- \rightarrow \rm hadrons$.  In figs. 3 - 6, 
the lower bound on ($d,\; \tan\beta,\;M_2$), respectively is determined by 
the allowed contribution to $\Delta \Gamma_Z(\rm visible) \le 23.1 
\times 10^{-3}$ GeV\cite{visiblez}.  This provides a weak lower bound on 
neutralino and chargino masses.   A better bound can be obtained
by   $e^+ e^- \rightarrow  \rm hadrons$. 
\renewcommand{\thefigure}{7 (\alph{figure})}\setcounter{figure}{0}
\begin{figure}
	\centerline{\psfig{file=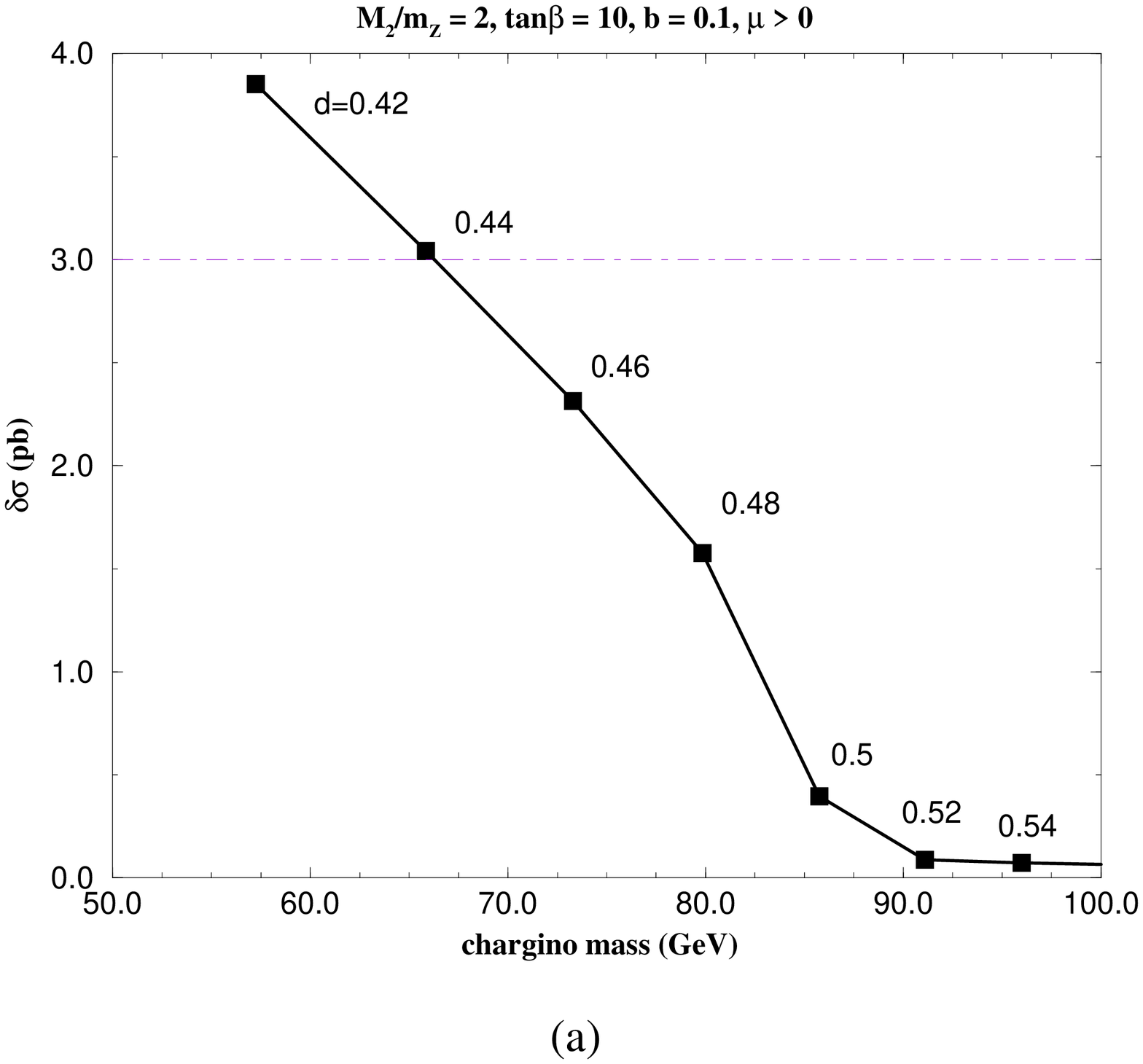,width=17cm,rheight=15.5cm,angle=-90}}
\caption{The cross section of $e^+e^- \rightarrow $ hadrons at
center of mass energy of 172 GeV from the chargino and neutralino pair 
production for the case $M_2/m_Z=2, \tan\beta=10, b=0.1$, and $\mu>0$ 
as a function of the chargino mass (varying $d$). We also show the OPAL 
2 $\sigma$ bound on new physics (dot-dashed line).}
\label{fig7a}
\end{figure}
\begin{figure}
	\centerline{\psfig{file=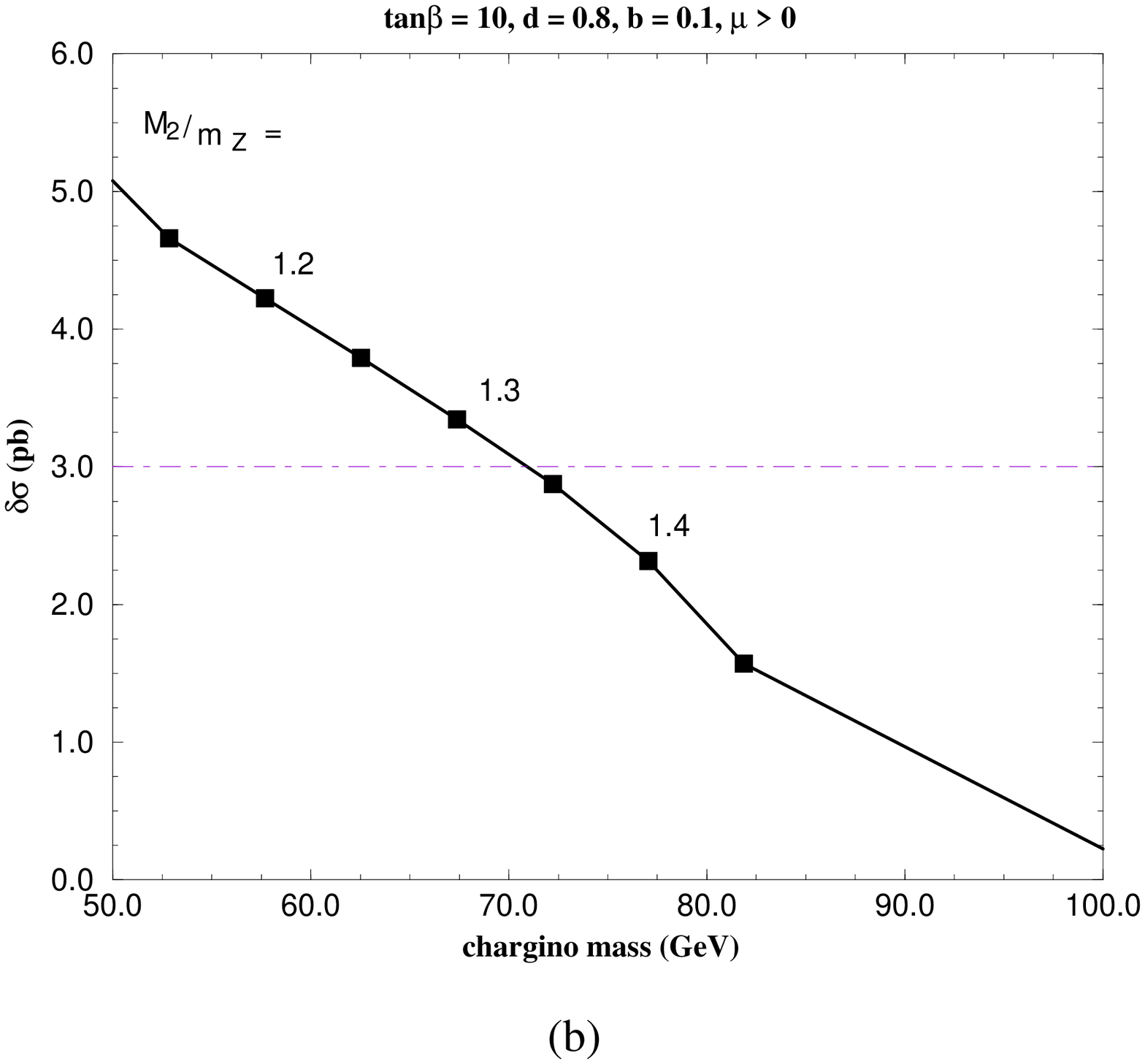,width=17cm,rheight=15.5cm,angle=-90}}
\caption{Same as fig.{\ref{fig7a}} for the case $\tan\beta=10, d=0.8, b=0.1$, 
and $\mu>0$ as a function of the chargino mass (varying $M_2$).}
\label{fig7b}
\end{figure}
In figs. 7(a,b) we give the 
constraints on the chargino mass coming from the process $e^+ e^- \rightarrow$ 
hadrons.\footnote{Here we assume that the gluino produces a visible jet.}
The horizontal line is the OPAL 2 $\sigma$ bound on new physics at 
$\sqrt{s} = 172$ GeV\cite{opal}.  In fig.  7(a) we show the contribution to 
$e^+ e^- \rightarrow  \rm hadrons$ due to chargino and neutralino production 
times the branching fraction for quark decay as a function of the chargino
mass (varying $d$) for fixed $\tan\beta = 10, M_2/m_Z = 2, \mu > 0, b = 0.1$. 
  We find a 
lower bound $d \ge 0.44$.  This translates into a lower bound on the lightest 
chargino mass $m_{\tilde \chi_1^+} \ge 66$ GeV (fig. 7(a)) and neutralino mass 
$m_{\tilde \chi_1^0} \ge 34$ GeV (fig. 3).  In fig. 7(b) we show the same 
constraints now varying $M_2$ instead of $d$.  For $d = 0.8$ and $M_2 > 122$ 
GeV, we find $m_{\tilde \chi_1^+} > 71$ GeV and  $m_{\tilde \chi_1^0} > 27$ GeV.

An additional test from LEP is on the number of 4 jet events.
The dominant decay mode 
for down squarks is $\tilde d \rightarrow d \tilde g$.  In fig. 4(b) we
see that the bottom squark can be as light as 75 GeV for large $\tan\beta$.
Nevertheless the calculated cross section for $e^+e^- \rightarrow
\tilde b \tilde {\bar b}$ is significantly below the OPAL bound for
excess four jet events at $\sqrt{s} = 184$ GeV\cite{opal4j}. Four jets are also 
possible via the direct process   $e^+ e^- \rightarrow q \bar q \tilde g 
\tilde g$.  Baer et al.\cite{baer} find strong limits using OPAL and L3 data\cite{opal2} 
for $e^+e^- \rightarrow Z$ with $Z \rightarrow (2 \; \rm or \; 4)$ jets + 
missing energy.  This data was used to place limits on the CMSSM process
$e^+e^- \rightarrow \chi^0_1 \; \chi^0_2$ with the subsequent decay
$\chi^0_2 \rightarrow q \; \bar q \; \chi^0_1 $.  They argue that a gluino mass
in the range $ 5 \le m_{\tilde g} \le 25$ GeV is ruled out.   This result 
seems to 
be relatively insensitive to the probability $P$.  Results from LEP2 at
higher center of mass energies do not improve upon this limit.

\renewcommand{\thefigure}{8 (\alph{figure})}\setcounter{figure}{0}
\begin{figure}
	\centerline{\psfig{file=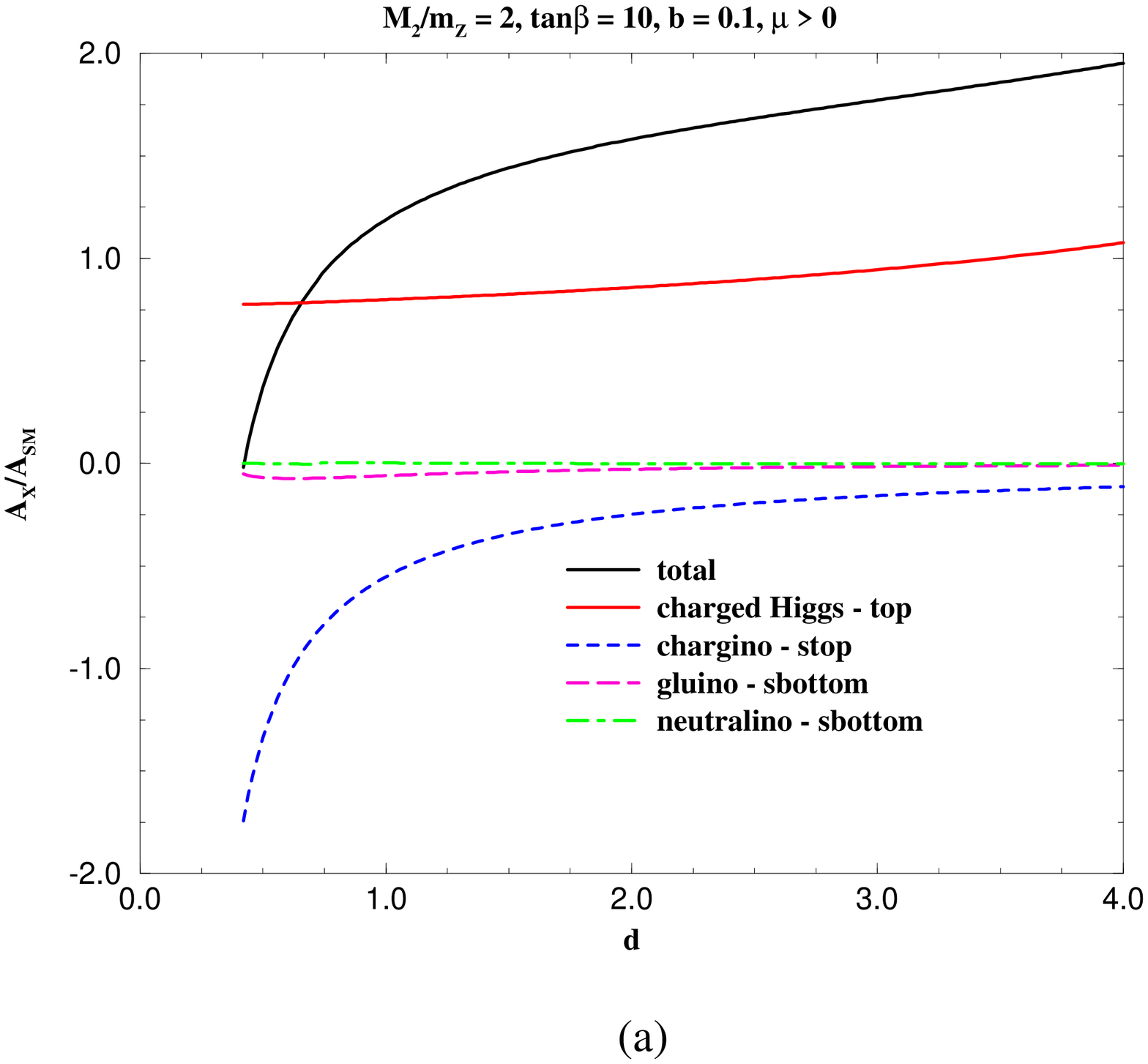,width=17cm,rheight=15cm,angle=-90}}
\caption{The ratio of $b \rightarrow s\gamma$ amplitude of SUSY
contribution to Standard Model one. The colored lines represent the
contribution of charged Higgs-top (red), chargino-stop (blue dashed),
gluino-sbottom (pink long dashed), neutralino-sbottom (green dot-dashed),
and total (all SUSY contribution plus SM contribution) (black solid line)
for $M_2/m_Z=2, \tan\beta=10, b=0.1$, and $\mu>0$ as a function of $d$.}
\label{fig8a}
\end{figure}
\begin{figure}
	\centerline{\psfig{file=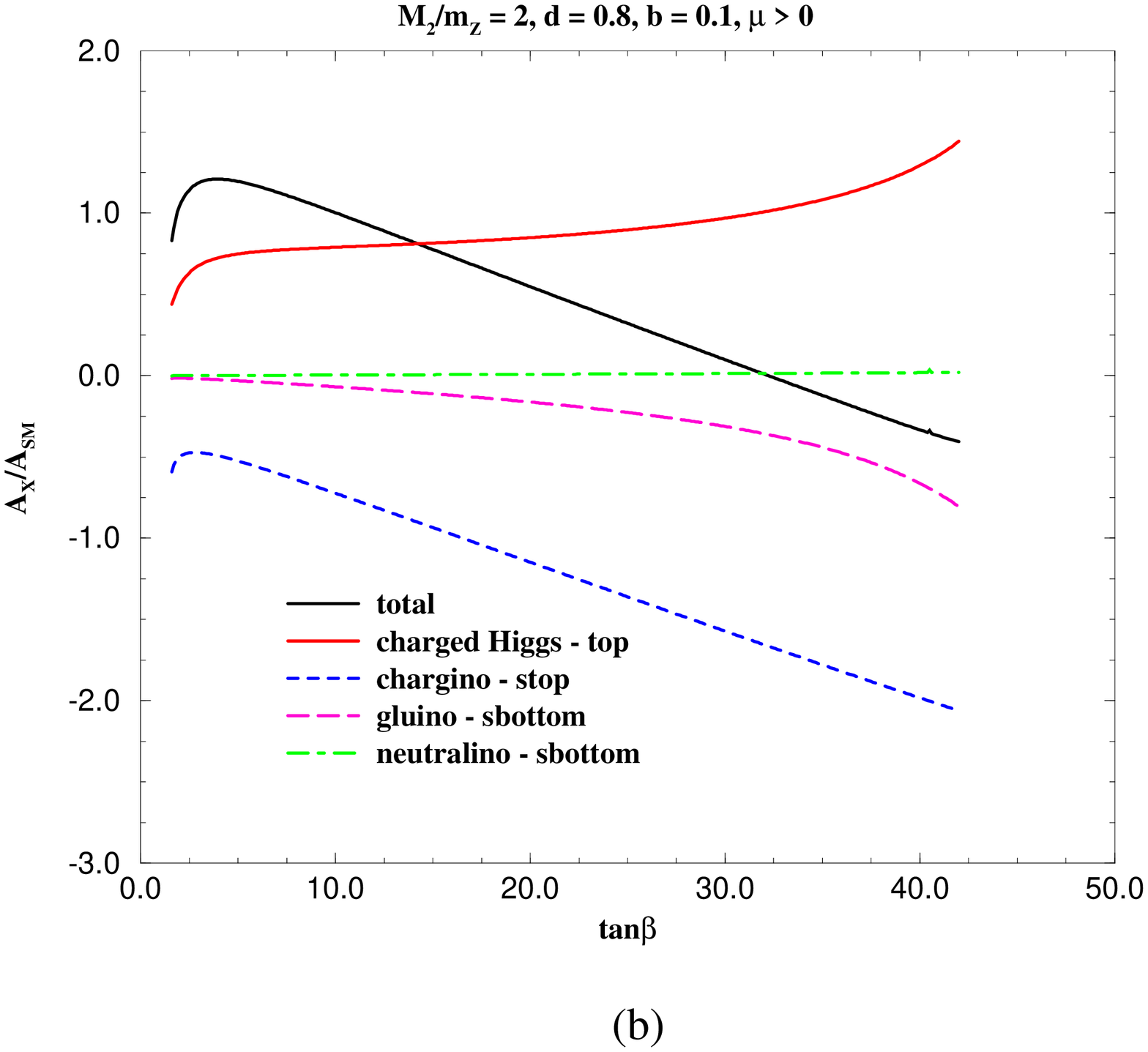,width=17cm,rheight=15.5cm,angle=-90}}
\caption{Same as (a) for $M_2/m_Z=2, d=0.8, b=0.1$, and $\mu>0$
as a function of $\tan\beta$.}
\label{fig8b}
\end{figure}
\begin{figure}
	\centerline{\psfig{file=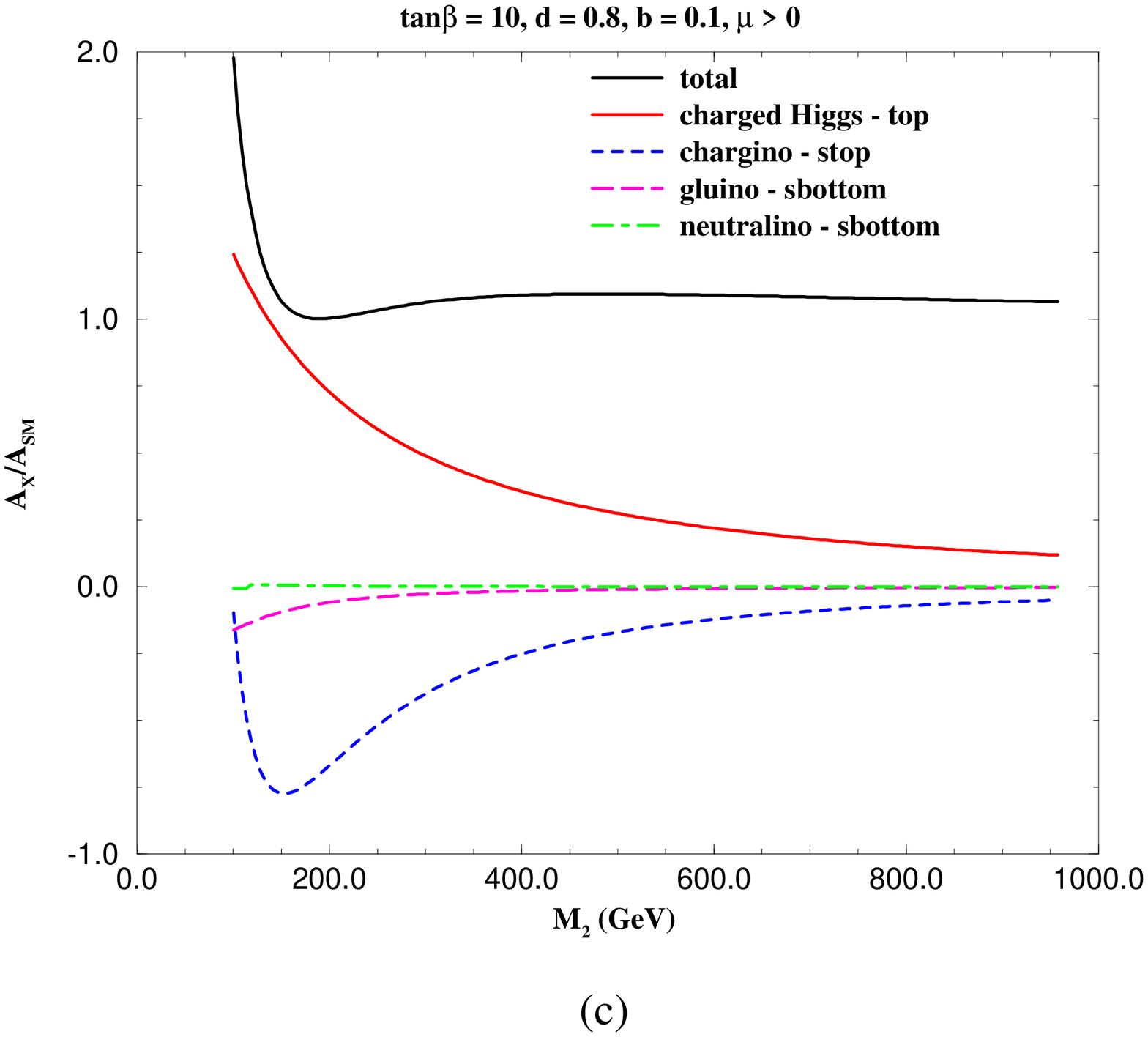,width=17cm,rheight=15.5cm,angle=-90}}
\caption{Same as (a) for $\tan\beta=10, d=0.8, b=0.1$, and $\mu>0$
as a function of $M_2$.}
\label{fig8c}
\end{figure}
\begin{figure}
	\centerline{\psfig{file=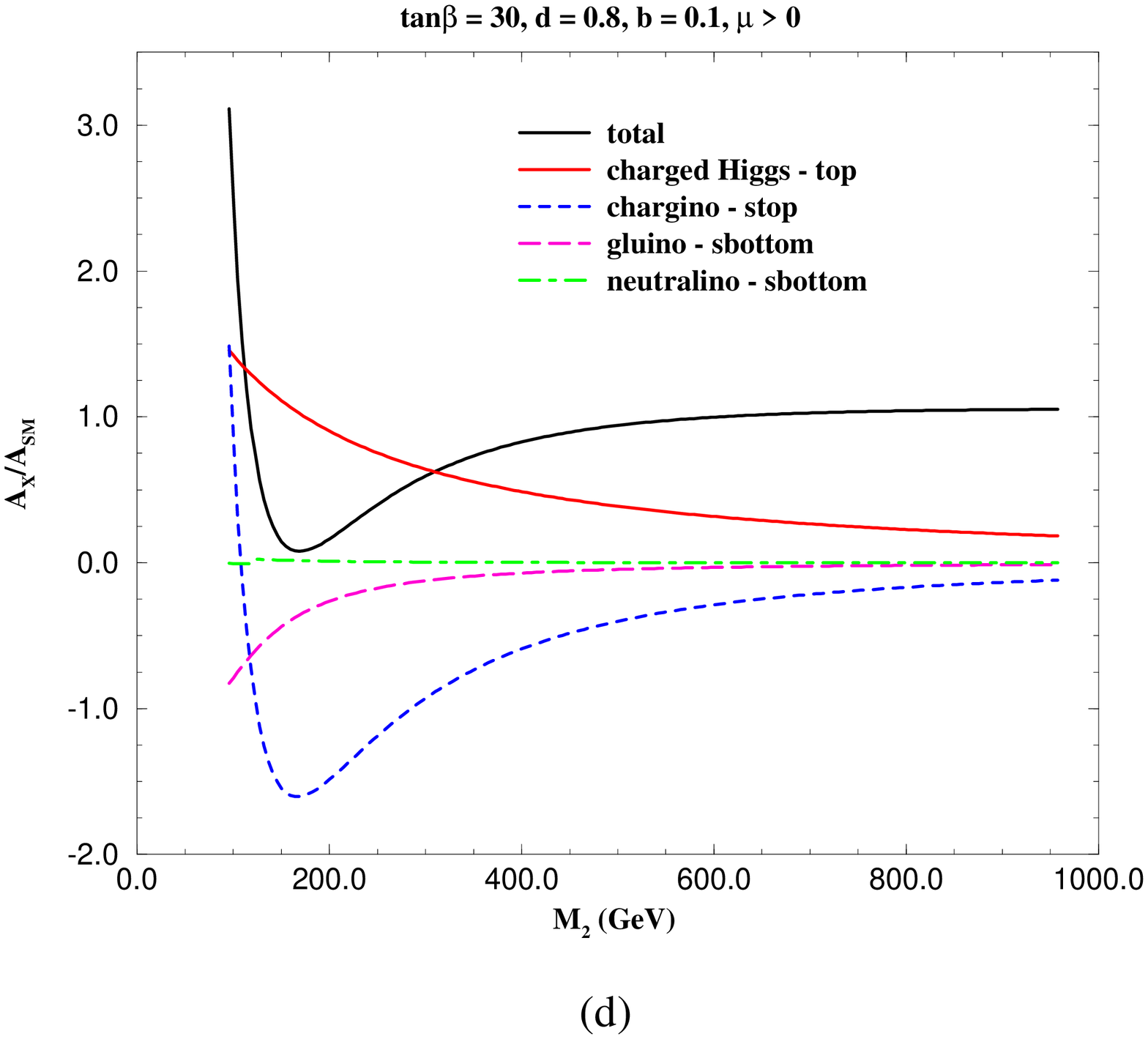,width=17cm,rheight=15.5cm,angle=-90}}
\caption{Same as (a) for $\tan\beta=30, d=0.8, b=0.1$, and $\mu>0$
as a function of $M_2$.}
\label{fig8d}
\end{figure}

We have evaluated the rate for $b \rightarrow s\gamma$.  The total contribution 
for $\mu < 0$ is always above the SM result, approaching it asymptotically 
for large $M_2$.\footnote{This is the reason figs. 3 - 6 are given only for 
$\mu > 0$.}   The results for $\mu > 0$ are given in figs. 8(a,b,c,d).  It 
is clear that there is a large range of parameters for which the total 
contribution is below the SM value.  Our result however only includes one 
loop analysis and should be taken only as an indication that it is
possible to obtain results consistent with the data\cite{bsgamma}.

For $d < 0.7$ (fig. 3) or $\tan\beta < 3.8$ (fig. 4a) or for any value of
$\tan\beta$ and $d = 0.5$ (fig. 4b), we have $\tilde
m_{t_1} < m_t$.  Then for  a sufficiently light gluino, the top can decay into a stop +
gluino.  This  range of parameters is testable at the Tevatron.  In the case that the stop 
is heavier than the lightest chargino, the dominant decay mode of the stop is 
$\tilde t_1 \rightarrow b \tilde \chi_1^+$.   In this case stop decay can 
mimic a top signal.  We are now studying this situation.  

It has also been argued by Baer et al.\cite{baer} that CDF data for jets + missing 
energy places stringent limits on the gluino mass.    These results however are
sensitive to the parameter $P$.   For $P \le 1/2$ they rule out a gluino with
mass in the range  $20 \lsim m_{\tilde g} \lsim 140$ GeV.  However for $P = 3/4$
there is an allowed window for the gluino mass  $25 \lsim m_{\tilde g} \lsim 35$
GeV.   Clearly these results are more sensitive to the details of how an
$R_0$ or $\tilde \rho^+$ interacts in the detector.   For example,
as discussed in \cite{baer},  a $\tilde \rho^+$ has some probability of being 
mis-identified as a muon if it is seen in the muon tracking chambers.   It 
is important that this analysis be redone by both CDF and D0.   In addition, 
Baer et al. have explicitly assumed that squarks are very heavy and hence do 
not contribute to the jet + missing energy signal.  Inclusion
of squarks can, via coherent interference, decrease the gluino production 
cross section and hence the rate for jets + missing energy.  At the same time, 
squark production and subsequent decay can increase the jets + missing energy 
rate.  Hence the Baer et al. analysis must be redone for the model presented
in this paper.

 \section{Conclusions}
In this paper we have presented a phenomenologically acceptable model with a 
gluino or gravitino LSP.  In case the gravitino is the LSP, then the gluino 
is the NLSP.  It can decay but only on a cosmological time scale.  Thus in 
either case the gluino is stable with respect to any accelerator experiment 
searching for SUSY.  We have also discussed some experimental bounds and 
possible tests of the theory at LEP, CLEO and the Tevatron.
We believe that further experimental studies are now justified.  

The theory includes a combination of gauge mediated and D term SUSY breaking.  
The latter contribution was shown to be necessary in order to obtain a Higgs 
with mass consistent with present experimental bounds.  

The gluino is naturally light in this theory because the {\bf 10} of Higgs 
(in SO(10)) mixes with the {\bf 10} of messengers.   It is light 
because of two effects.  Firstly, the SU(3) triplet messengers are heavier 
than the electroweak doublet messengers; suppressing SUSY breaking mediated by 
color interactions.  Secondly, R symmetry (which keeps gauginos massless) is 
broken at the electroweak doublet messenger scale; providing an additional 
suppression.

There are several other consequences of a theory with a gluino LSP which may 
have cosmological and/or astrophysical significance.   It has been shown that
electroweak baryogenesis is not possible in the SM.  On the other hand, the
necessary conditions for electroweak baryogenesis in the MSSM requires
$m_{h} \sim 100$ GeV and $\tilde m_{t_R}^2 < 0$\cite{carena}.  
{\em These conditions are naturally satisfied in our model.}  They gaurantee that there 
is a first order phase transition which is sufficiently strong to shut off
baryon violating interactions inside the bubbles of broken electroweak  phase. 
If there is also sufficient CP violation; then baryogenesis is possible.   
  
In addition, the $R_0$ provides a natural particle physics candidate for 
the source of ultra high energy cosmic rays, the UHECRON.\footnote{This
is even possible if the gluino is the NLSP and decays into a gravitino LSP.
For example, a gluino with mass $\sim 25$ GeV and energy characteristic of
the highest energy cosmic rays $\sim 10^8$ TeV can travel $\sim 10^5$ Mpc
before it decays.}  It has recently 
been argued\cite{kolb} that an $R_0$ with mass in the range from 2 - 50 GeV
could reproduce the highest energy cosmic ray shower observed by the Fly's Eye 
Detector\cite{fly}.

Finally, the $R_0$ LSP is not a dark matter candidate since the annihilation 
cross section is too large and hence the relative abundance of $R_0$ to 
baryons is too small.\footnote{The uncertainty in the annihilation cross 
section is quite large $\sim 10^{\pm 1.5}$.  Nevertheless the relative 
abundance of gluinos to baryons is of order $(10^{-10} - 10^{-7}) 
(m_{\tilde g}/1 \rm GeV)$.  For a recent re-analysis,see Baer et al.\cite{baer}}
  There are nevertheless severe constraints
on the abundance of $R_0$s coming from searches for anomalous heavy isotopes 
of hydrogen and oxygen\cite{isotope}.   If $R_0$ binds to hydrogen, then any 
$R_0$ with mass greater than 2 GeV is ruled out.   In addition, if the 
concentration of $R_0$ in oxygen is greater than one part in $10^{16} - 
10^{19}$, it is also ruled out.   Note however, if the gravitino is the LSP 
and $R_0$ is the NLSP, these limits are evaded.

{\bf Acknowledgements}
Finally, this work is partially supported by DOE grant DOE/ER/01545-743. 
We would like to thank A. Mafi, S. Mrenna and U. Sarid for useful contributions and
K. Cheung, G. Farrar, J. Gunion, and E.W. Kolb for discussions.
\medskip

%

\newpage

\begin{thebibliography}{99}
\bibitem{sugra} A.H. Chamseddine, R. Arnowitt and P. Nath, 
{\it Phys. Rev. Lett.}{\bf 29}, 970 (1982);  R. Barbieri, S. Ferrara and 
C.A. Savoy, {\it Phys. Lett.}{\bf B119}, 343 (1983);   L.J. Hall, J. Lykken 
and S. Weinberg, {\it Phys. Rev.}{\bf D22}, 2359 (1983);  P. Nath, R. Arnowitt 
and A.H. Chamseddine, {\it Nucl. Phys.} {\bf B322}, 121 (1983).
%
\bibitem{cmssm} G.L. Kane, C. Kolda, L. Roszkowski and J.D. Wells, 
{\it Phys. Rev.} {\bf D49}, 6173 (1994).
%
\bibitem{gmsb} M. Dine, W. Fischler, and M. Srednicki,  {\it Nucl. Phys.}  
{\bf B189}, 575 (1981); S. Dimopoulos and S. Raby, {\it Nucl. Phys.}{\bf B192}, 
 353 (1981); M. Dine and W. Fischler, {\it Phys. Lett.}{\bf B110}, 227 (1982);
M. Dine and M. Srednicki, {\it Nucl. Phys.}{\bf B202},  238 (1982);
L. Alvarez-Gaum\'{e}, M. Claudson, and M. Wise, {\it Nucl. Phys.}{\bf  B207}, 96
(1982); C. Nappi and B. Ovrut,  {\it Phys. Lett.}{\bf  B113}, 175 (1982).
\bibitem{dine} M. Dine, A.E. Nelson and Y. Shirman, {\it Phys. Rev.}{\bf  D51},
1362  (1995); M. Dine, A.E. Nelson, Y. Nir and Y. Shirman,
{\it Phys. Rev.}{\bf D53},  2658 (1996). 
%
\bibitem{raby} S. Raby, {\it Phys. Rev.} {\bf D56}, 2852 (1997) and 
{\it Phys. Lett.} {\bf B422}, 158 (1998).	
%
\bibitem{tobe}  S. Raby and K. Tobe, "Dynamical SUSY Breaking with
a Hybrid Messenger Sector," OHSTPY-HEP-T-98-009, (hep-ph/9805317).
%
\bibitem{fidterm}  P. Fayet and J. Iliopoulos, {\it Phys. Lett.}{\bf B51}, 461
(1974) ; W. Fischler, H.P. Nilles, J. Polchinski, S.
Raby and L. Susskind,  {\it Phys. Rev.  Lett.}{\bf 47}, 757-759 (1981);
 M. Dine, N. Seiberg   and E. Witten,  {\it Nucl.  Phys.}
{\bf B289}, 589 (1987) ;  J. Atick,  L. Dixon  and A. Sen,  {\it Nucl.  Phys.}
{\bf B292}, 109 (1987); M. Dine, I. Ichinose and N. Seiberg,  {\it Nucl. 
Phys.} {\bf B293}, 253 (1988).
%
\bibitem{horava}  P. Horava and E. Witten,  {\it Nucl.  Phys.}
{\bf B460}, 506 (1996).
%
\bibitem{jmr} J. March-Russell, ``The Fayet-Iliopoulos term in Type I
string theory and M-theory," (hep-ph/9806426).
%
\bibitem{dw} Dimopoulos and Wilczek, Preprint NSF-ITP-82-07 (1982).
%
\bibitem{babu} K.S. Babu and S. Barr, {\it Phys. Rev.} {\bf D48},
5354 (1993).	
%
\bibitem{farragi} A.E. Faraggi and J.C. Pati, "A Family-Universal Anomalous
U(1) in String Models as the Origin of Supersymmetry Breaking and Squark
Degeneracy," (hep-ph/9712516).
%
\bibitem{giudice} G.F. Giudice and R. Rattazzi, {\it Nucl. Phys.}
{\bf B511}, 25 (1998).
%
\bibitem{hall} L.J. Hall, R. Rattazzi, and U. Sarid, {\it Phys. Rev.}
{\bf D50}, 7048 (1994); M. Carena, M. Olechowski, S. Pokorski and
C. Wagner, {\it Nucl. Phys.}{\bf B426} 269 (1994);
R. Hempfling,  {\it Z. Phys.} {\bf C63} 309 (1994).
%
\bibitem{Higgsbound} L3 Collaboration (M. Acciarri et al.), CERN-EP-98-072;
OPAL Collaboration, CERN-EP/98-093; ALEPH  Collaboration, 
ALEPH 98-029 (CONF 98-017).
%
\bibitem{charginobound}DELPHI Collaboration (P. Abreu et al.),
		{\it Eur. Phys. J.} {\bf C1}, 1 (1998);
	OPAL Collaboration (K. Ackerstaff et al.),
		{\it Eur. Phys. J.} {\bf C2}, 213 (1998);
	ALEPH Collaboration (R. Barate et al.),
		{\it Eur. Phys. J.} {\bf C2}, 3417 (1998);
	L3 Collaboration (M. Acciarri et al.), CERN-PPE-97-130.
%
\bibitem{baer} H. Baer, K. Cheung and J.F. Gunion, "A Heavy Gluino
as the Lightest Supersymmetric Particle," (hep-ph/9806361).
%
\bibitem{visiblez} L3
Collaboration (M. Acciarri et al.),
		{\it Phys. Lett.} {\bf B350}, 109 (1995).
%
\bibitem{opal} OPAL Collaboration (K. Ackerstaff et al.),
		{\it Eur. Phys. J.}{\bf C2}, 441 (1998).
%
\bibitem{opal4j} OPAL Collaboration, CERN-EP/98-013, (hep-ex/9802015).	
%
\bibitem{opal2} OPAL collaboration (K. Ackerstaff et al.),
		{\it Phys. Lett.} {\bf B377}, 273 (1996);
		L3 Collaboration (M. Acciarri et al.),
		{\it Phys. Lett.} {\bf B350}, 109 (1995).
%
\bibitem{bsgamma} CLEO collaboration (M.S. Alam et al.),
		{\it Phys. Rev. Lett.} {\bf 74}, 2885 (1995).
%
\bibitem{carena} M. Carena, M. Quiros and C.E.M. Wagner, 
{\it Phys. Lett.} {\bf B380}, 81 (1996).
%
\bibitem{kolb} D.J.H. Chung, G.R. Farrar and E.W. Kolb,
{\it Phys. Rev.} {\bf D57}, 4606 (1998).
%
\bibitem{fly} D.J. Bird et al., {\it Astrophys. J.} {\bf 424}, 491 (1994).
%
\bibitem{isotope} R.A. Muller et al., {\it Science} (April 29, 1977), 521;
P.F. Smith and J.R.J. Bennett, {\it Nucl. Phys.} {\bf B149}, 525 (1979);
P.F. Smith et al., {\it Nucl. Phys.} {\bf B206}, 333 (1982).
\end{thebibliography}
\end{document}